\documentclass[a4paper,fleqn,usenatbib]{mnras}

\usepackage[T1]{fontenc}
\usepackage{ae,aecompl}

\usepackage{graphicx}	
\usepackage{amsmath}	
\usepackage{amssymb}	
\usepackage{bm}		
\usepackage{color}

\usepackage{txfonts} 
\usepackage{hyperref}

\newcommand{\HII}{\ion{H}{ii}}
\newcommand{\HI}{\ion{H}{i}}
\newcommand{\SII}{[\ion{S}{ii}]}
\newcommand{\NII}{[\ion{N}{ii}]}
\newcommand{\OIII}{[\ion{O}{iii}]}
\newcommand{\OI}{[\ion{O}{i}]}
\newcommand{\Ha}{H$\alpha$}
\newcommand{\Hb}{H$\beta$}
\newcommand{\SN}{S$/$N}

\title[Galactic winds in UGC 10043]{Star formation driven galactic winds in UGC 10043}

\author[C. L\'opez-Cob\'a et al]{C. L\'opez-Cob\'a$^{1}$,
S.~F. S\'anchez$^{1}$,
A.~V. Moiseev$^{2}$,
D.~V. Oparin$^{2}$,
T. Bitsakis$^{3},$\newauthor
I. Cruz-Gonz\'alez$^{1}$,
C. Morisset$^{1}$,
L. Galbany$^{4,5}$,
J. Bland-Hawthorn$^{6}$,\newauthor
M.~M. Roth$^{7}$,
R.-J. Dettmar$^{8}$,
D.~J. Bomans$^{8}$,
Rosa M. Gonz\'alez Delgado$^{9}$,
M. Cano-D\'{\i}az$^{10}$,\newauthor
R.~A. Marino$^{11}$,
C. Kehrig$^{12}$,
A. Monreal Ibero$^{13}$
and V. Abril-Melgarejo$^{14}$
\\
$^{1}$Instituto de Astronom\'ia, Universidad Nacional Aut\'onoma de  M\'exico, A.~P. 70-264, C.P. 04510, M\'exico, D.F., Mexico \\
$^{2}$Special Astrophysical Observatory, Russian Academy of Sciences, Nizhnii Arkhyz 369167, Russia\\
$^{3}$CONACYT Research Fellow - Instituto de Radioastronom\'ia y Astrof\'isica, Universidad Nacional Aut\'onoma de M\'exico, C.P. 58190, Morelia, Mexico\\
$^{4}$Pittsburgh Particle Physics, Astrophysics, and Cosmology Center (PITT PACC), USA.\\
$^{5}$Physics and Astronomy Department, University of Pittsburgh, Pittsburgh, PA 15260, USA.\\
$^{6}$Sydney Institute for Astronomy, School of Physics, University of Sydney, NSW 2006, Australia\\
$^{7}$Leibniz-Institut f\"ur Astrophysik Potsdam (AIP), An der Sternwarte 16, D-14482 Potsdam, Germany\\
$^{8}$Astronomisches Institut, Ruhr-Universit{\"a}t Bochum, Universit{\"a}tsstr. 150, D-44801 Bochum, Germany\\
$^{9}$ Instituto de Astrof\'isica de Andaluc\'ia (IAA/CSIC), Glorieta de la Astronom\'{\i}a s/n Aptdo. 3004, E-18080 Granada, Spain\\
$^{10}$ CONACYT Research Fellow - Instituto de Astronom\'ia, Universidad Nacional Aut\'onoma de M\'exico, Apartado Postal 70-264, M\'exico D.F., 04510 Mexico\\
$^{11}$Department of Physics, Institute for Astronomy, ETH Z$\ddot{u}$rich, CH-8093 Z$\ddot{u}$rich, Switzerland\\
$^{12}$Instituto de Astrof\'isica de Andaluc\'ia (CSIC), Glorieta de la Astronom\'ia s/n Aptdo. 3004, 18080, Granada, Spain\\
$^{13}$GEPI, Observatoire de Paris, PSL Research University, CNRS, Universit\'e Paris-Diderot, Sorbonne Paris Cit\'e, Place Jules Janssen,
92195 Meudon, France\\
$^{14}$Departamento de F\'isica, Universidad de los Andes, Cra. 1 No. 18A--10, Edificio Ip, A. A. 4976 Bogot\'a, Colombia
}

\date{Accepted XXX. Received YYY; in original form ZZZ}

\pubyear{2016}

\begin{document}
\label{firstpage}
\pagerange{\pageref{firstpage}--\pageref{lastpage}}
\maketitle

\begin{abstract}

We study the galactic wind in the edge-on spiral galaxy UGC 10043 with
the combination of the CALIFA integral field spectroscopy data, scanning Fabry-Perot interferometry (FPI), and multiband photometry. We detect ionized gas in the extraplanar regions
reaching a relatively high distance, up to $\sim$ 4 kpc above the galactic disk. The ionized gas line ratios (\NII/\Ha, \SII/\Ha~and \OI/\Ha) present an enhancement along the semi minor axis, in contrast with the values found at the disk, where they are compatible with ionization due to \HII-regions. These differences, together with the biconic symmetry of the extra-planar ionized structure, makes UGC 10043 a clear candidate for a galaxy with gas outflows ionizated by shocks.
From the comparison of shock models with the observed line ratios, and the kinematics observed from the FPI data, we constrain the physical properties of the observed outflow. The data are compatible with a velocity increase of the gas along the extraplanar distances up to $<$ 400 km s$^{-1}$ and the preshock  density decreasing in the same direction. We also observe a discrepancy in the SFR estimated based on \Ha~ (0.36 M$_{\sun}$ yr$^{-1}$) and the estimated with the CIGALE code, being the latter 5 times larger. Nevertheless, this SFR is still not enough to drive the observed galactic wind if we do not take into account the filling factor. We stress that the combination of the three techniques of observation with models is a powerful tool to explore galactic winds in the Local Universe.
\end{abstract}




\section{Introduction}

Galactic winds driven by a combination of supernovae explosions
and stellar winds from massive stars have been proposed as a regulating 
mechanism in the formation and evolution of galaxies \citep[e.g.][]{Silk1998,Springel,Ciardi}. 
They can also explain the origin
of some global properties of galaxies such as (i) the 
mass metallicity relation \citep[e.g.][]{Tremonti2004, Finlator, Peeples},
(ii) the shape of the stellar mass function \citep[e.g.][]{Nakano,Elmegreen} and (iii) the enrichment of
the intergalactic medium with metals \citep[e.g.][]{Aguirre2001,Oppenheimer}.

Since the discovery of the central-starburst driven wind in M 82
\citep{Lynds}, there have been several works aiming to understand the
process that drive galactic winds in active galactic nuclei (AGN),
starburst and merging galaxies
\citep{Heckman,Rich2010,Rich2011,Sturm,Wild}. The understanding of
feedback processes by outflows plays an important role in the study of
galaxy evolution since they may affect strongly the properties of the
interestellar medium: i.e. the energy released by star formation (SF, hereafter), and AGN
activity into the halo can heat the gas, preventing its cooling and
as a consequence suppressing the SF in low-mass galaxies
\citep[e.g.][]{Scannapieco,Hopkins2012}.  Such processes may have a
significant impact on the evolution of the host galaxy by regulating
the amount of cold gas available for SF. In addition, the
galactic winds can reduce the number of existing dwarf galaxies, since 
the kinetic energy from supernovae ejects halo gas, thus suppressing the SF \citep[e.g.][]{Dekel}.

\begin{figure}
 \centering
  \includegraphics[clip=true,trim=0cm 0cm 0cm 0cm,width=\columnwidth]{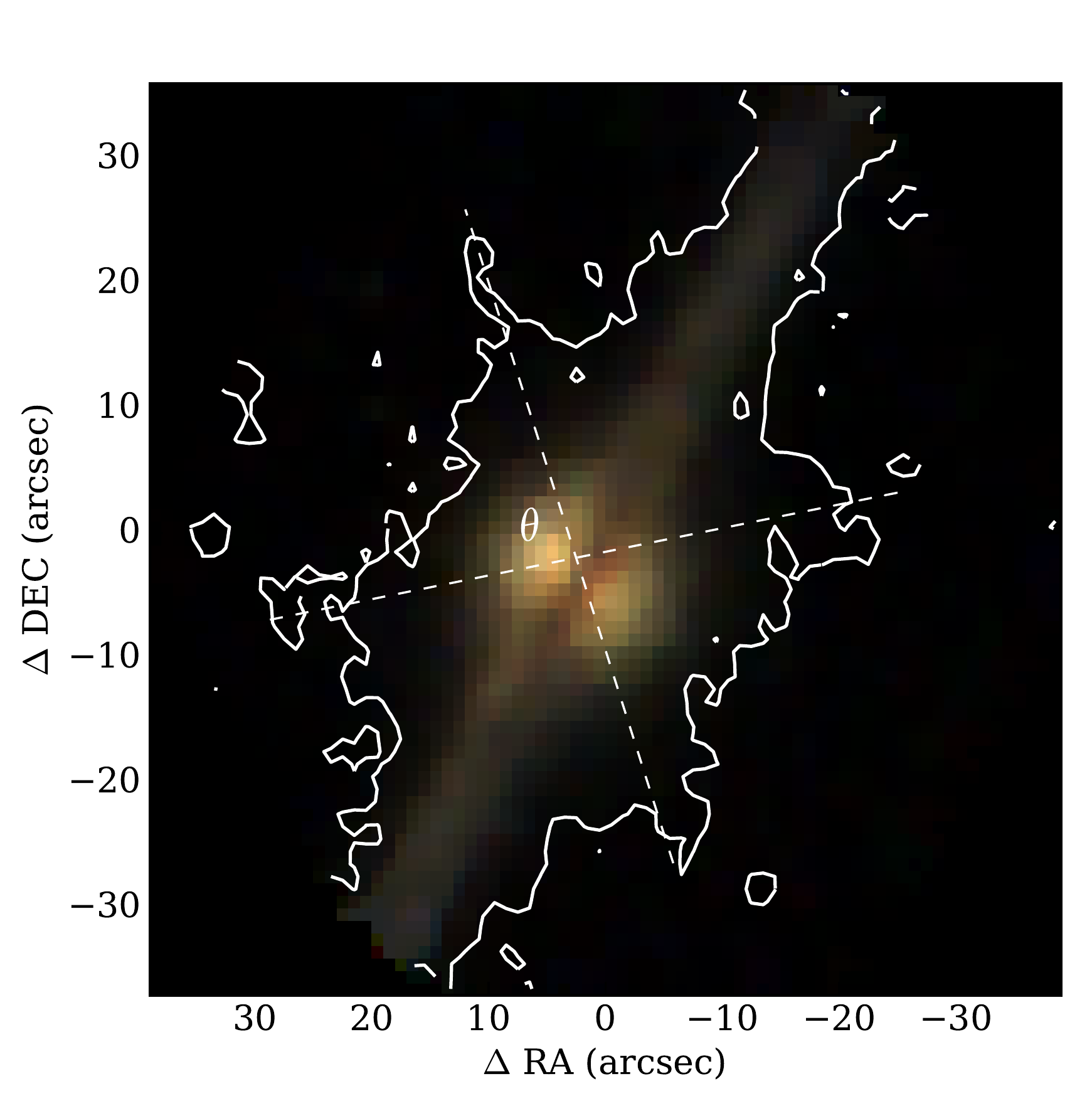}
 \caption{True-color image of UGC 10043 extracted from the CALIFA data
   cube, where blue corresponds to 4450 \AA, green to 5500 \AA~ and red to 6580 \AA. It 
is appreciated the dust absorption line along the galactic
   disk, that is more evident in the central regions. The white contour
   indicates the detection limit of the flux intensity of the
   \NII$\lambda$6584 emission line. The dashed lines indicate the potential bi-conical
 structure of the ionized gas emission, with an aperture
   angle of $\theta\approx$ 80\degr traced by eye at the maximum extension of the emission line intensity. }
 \label{rgb}
\end{figure}

Super-winds are phenomenologically complex since they are comprised of
multiple gas phases moving at different velocities
\citep[e.g.][]{Bland,Heckman2000,Veilleux2005,Strickland}. The mass
released by starburst superwinds can range from
tens to thousands of M$_{\sun}$ yr$^{-1}$ with velocities of $\sim$
100 km s$^{-1}$ for the cool component, meanwhile in AGN-driven outflows the
velocities exceeds  500 km s$^{-1}$.  The mechanism
that drives outflows in starbursts is the mechanical energy supplied
by supernovae explosions and stellar winds \citep[e.g.][]{Leitherer}.
It is known
that super-winds at high scales are very common in galaxies with 
SF surface density larger than 0.1 M$_{\odot}$ yr$^{-1}$
kpc$^{-2}$, both in the local Universe \citep{Dahlem, Lehnert} and at
high redshift \citep{Pettini}.

Starburst galaxies are commonly observed to host supernovae-driven
galactic scale winds \citep[e.g.][]{Heckman,Heckman2003}.
A galactic wind is produced when the kinetic energy ejected by
supernovae and stellar winds, produced by massive star formation, is
efficiently thermalized. This means that the kinetic energy of the
wind is converted into thermal energy via shocks, with a little loss
of energy by radiation due to the high temperatures and low
densities. The joint effect of supernovae and stellar winds creates a
bubble of hot gas (T $\sim10^8$ K) inside the star-forming region with a
pressure greater than its environment \citep[e.g.][]{Heckman}.

As the bubble expands and sweeps its environment, it quickly turns
into a radiative phase.  If the environment is stratified, like in
the case of galactic disks, the bubble will expand faster in the vertical
direction of the pressure gradient. The velocity of this hot gas is expected to be in the range of a few thousands km s$^{-1}$ \citep[e.g.][]{Heckman2003}. Once the bubble reaches 
a certain scale height, the expansion will accelerate and break into fragments. This
will allow the gas to expand freely into the halo and the intergalactic
medium  follows a bipolar collimated outflow geometry. This is the so-called
blow-out phase when the bubble turns into a super-wind.  The optical
and X-ray emission in the blow-out phase comes from obstacles (clouds,
fragmented shells) that are immersed in the gas and are shock-heated by the outflow.
This scenario was explained in detail by \citet{Heckman}.

\begin{table}
  \caption{Parameters of UGC 10043. 
  References: (1) NASA/IPAC Extragalactic Database; 
  (2)  HyperLeda; (3) \citet{Matthews} } \label{tab:referencias}
  \begin{tabular}{llc}
    \hline
    Parameter & UGC~10043 &References \\
    \hline
    Morphology &Sbc&1\\
    $z$&0.007208&1\\
    $\alpha$ (J2000)&15h48m41.2s&1\\
    $\delta$ (J2000)&+21d52m09.8s &1\\
    Distance [Mpc]&34.995&2\\
    $m-M$ [mag]&32.72&2\\
    $P.A.$ [deg] &151.5&3 \\
    $i$ [deg]&90&3\\
    $h_z$ [pc]&395&3\\
    $L_\mathrm{H\alpha}$[erg s$^{-1}$]&4.5$\times$10$^{40}$&This paper\\
    log(M$_*$/M$_{\sun}$)&9.79&This paper\\
    \hline
    \end{tabular}
\end{table}

Huge advantages over classical long slit spectroscopy are obtained
by the use of
Integral Field
Spectroscopy (IFS) to study the spatially resolved properties of
galaxies. IFS provides with spatially resolved spectra of a complete
Field-of-View (FoV), allowing the study of different components of a
galaxy in a simultaneous way. The combination of image and
spectroscopy through IFS provides a better understanding of the
properties of galaxies.  Several works have implemented the use of
this technique to study the ionization by shocks and/or outflows of
galactic winds
\citep[e.g.][]{Rich2010,Rich2011,Rich2014,Ho2014,Wild,Bik,
  Mahony}. However, to our knowledge, there are no detailed studies of the frequency of
outflows in galaxies \citep[see for
  example][]{Dahlem,Heckman2000,Heckman2003,Ho2016}.

In our current paper, we present the study of the spatially resolved
properties of the ionized gas in UGC 10043 based on the data
from the Calar-Alto Legacy Integral Field Area (CALIFA) survey \citep{sanchez2012}, 
focusing on the analysis of the
detected galactic wind. This is a pilot study which ultimate goal is 
the exploration of the frequency and physical conditions of such outflows
in the complete sample of CALIFA. The paper is organized as
follows: we present the main properties of UGC 10043 in \S 2; the
data are described in \S 3, and the conducted analysis in
\S 4. Section 5 describes the  analysis of observations with a scanning 
Fabry-Perot interferometer at the Russian 6-m telescope; while the main results are presented in \S 6 . Finally, we present the
main conclusions and discussion in \S 7.  Throughout the paper, we adopted the
standard LCDM cosmology, with parameters $H_0=$ 70.4 km s$^{-1}$
Mpc$^{-1}$, $\Omega_\mathrm{m}$ = 0.268 and $\Omega_{\Lambda}$ =
0.732.

\begin{figure*}
 \includegraphics[clip=true,trim=2cm 0cm 0cm 2cm,width=\textwidth,height=!]{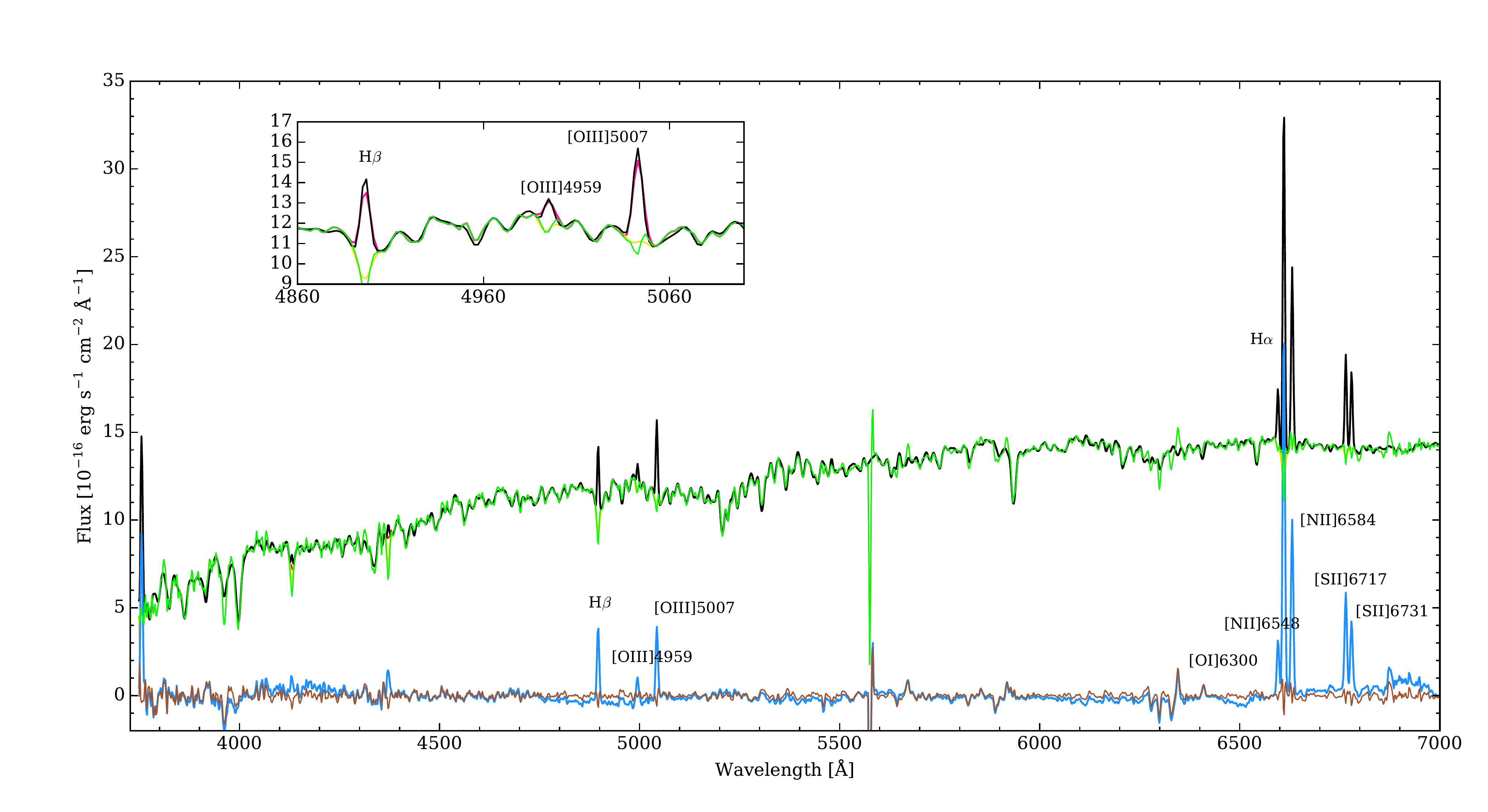}
 \caption{Example of the fitting procedure of the stellar population using \textsc{FIT3D}
 for one spectrum derived by integrating 25 spaxels around the central region of UGC 10043.
 The original spectrum is shown in black and in yellow is the best fit of the SSP. The 
 red correspond to the best combination of the stellar populations and emission lines 
 that describe the data. In green is
 the stellar population after the removal of the fitted emission lines. One of the results of 
 the fit is a spectrum without the stellar population including the emission lines
 of the ionized gas.
 In blue we present the emission lines after removing the fitted stellar population and in brown 
 the residual of the subtraction of the best fit including the stellar populations and the emission line
 model. The insight panel shows an enlargement around H$\beta$  and the oxygen doublet 
 [\ion{O}{III}] of the fit, the y axis has the same units in both plots.
}
  \label{ajuste}
\end{figure*}

\section{Main properties of UGC 10043}

UGC 10043 is an edge-on spiral galaxy ($i$ $\sim$ 90\degr) located at a
distance of $\sim$ 35 Mpc (see Table \ref{tab:referencias}). 
It is found close to MD2004 dwarf, a companion galaxy
located 84\arcsec to the NW. \citet{Matthews} suggested a possible interaction
with UGC 10043 which may explain the observed tidal warp in its disk.
UGC 10043 presents a very thin disk and a prominent large and bright
bulge (Fig. \ref{rgb}). Optical images reveal a very pronounced dust lane along its
disk.  \citet{Matthews}, using Hubble Space Telescope (HST) narrow band observations in
H$\alpha$ $+$ [\ion{N}{II}] found evidence of 
star-formation in its nucleus. They also found \Ha~ emission
perpendicular to the disk following an approximately bi-conic structure
resulting from a possible galactic wind. These authors estimate 
the velocity of the wind in $\sim$ 100 km s$^{-1}$ 
reaching a distance of 3.5 kpc over the disk.  \citet{Aguirre2009} mapped UGC 10043
with the VLA. They argued that UGC 10043 is under interaction with 
MCG $+$04$-$37$-$035 (located 2.5\arcmin to the W; z $=$ 0.007398)
evidenced by an observed \HI~ bridge between the two galaxies.

\section{Data}

In our work we are using the IFS observations of UGC10043 from the first and 
second CALIFA data releases \citep[e.g.][]{Husemann2013,Garcia-Benito}.
The CALIFA survey is a recently completed project 
that comprises three data releases (DR1, DR2 and DR3), the last one delivered in 2016 \citep{CALIFA_DR3}. 
The aim of CALIFA was to obtain spatially
resolved spectroscopy of more than 600 galaxies at the Local Universe
(0.005$<$ z $<$ 0.03) covering a wide range of morphological and stellar masses.
The observations cover the optical size of the
galaxies up to 2.5 effective radii using the wide-field Integral
Field Unit (IFU) Pmas fiber PAcK  \citep[PPAK;][]{Kelz} of the 
Potsdam Multi-Aperture Spectrophotometer instrument
\citep[PMAS;][]{Roth}.  The PPAK fiber bundle consists of 331 fibers of
2.7$''$ diameter each one covering a total hexagonal FoV of
$74''\times64''$ with a filling factor of $\sim 60\%$.  In order to
guarantee a complete coverage of the FoV, three dithering pointings
were applied to obtain 993 independent spectra for each object.  The
final spatial resolution is $\sim$1 kpc at the redshift of the
sample. This allows to resolve spatially the spectroscopic properties
from the most relevant components of the galaxies (\HII~ regions, bars, spiral arms, bulges).  Each galaxy of the
CALIFA sample was observed in two different configurations, one of low
resolution (V500, R$\sim$ 850) covering the optical range 3750--7500\,\AA~ and another one of intermediate resolution (V1200, R$\sim$ 1650)
that covers the blue part of the optical range of the spectra
3700--4800\,\AA. Along this article we use the data of the V500 set-up
for UGC 10043.

The data were reduced using version 1.5 of the CALIFA pipeline, whose
modifications with respect to the one presented in \citet{sanchez2012}
and \citet{Husemann2013}, are described in \citet{Garcia-Benito}.  The
procedure comprises the usual steps in reduction of IFS data, as
described in \citet{sanchez2006b}: bias and dark subtraction, cosmic-ray
removal, CCD flat-fielding, spectra tracing and extraction, wavelength
and flux calibration, and finally cube reconstruction. The final
product of the data reduction is a regular-grid data-cube, with $x$ and
$y$ coordinates indicating the right ascension and declination of
the target and $z$ a common step in wavelength. The CALIFA pipeline
also provides with the propagated error cube, a proper mask cube of bad
pixels, and a prescription of how to handle the errors when performing
spatial binning (due to covariance between adjacent pixels after image
reconstruction).

These data were complemented with multi-band, aperture matched,
photometry extracted from the Galaxy Evolution Explorer 
\citep[GALEX;][]{martin2015}, Sloan Digital Sky Survey \citep[SDSS;][]{york2000},
and Wide-Field Infrared Survey Explorer \citep[WISE;][]{wright2010} surveys,
extracted following the procedures
described in Bitsakis et al. (in prep.). 
To estimate the physical properties of this galaxy,
these authors provided a spectral energy distribution
(SED) modelling, using CIGALE \citep{Noll}.

We also use Fabry-P\'erot Interferometer
(FPI) observations obtained at the prime focus 
of the 6-m telescope of Special Astrophysical Observatory  Russian Academy of Sciences
(SAO RAS) in 2014 May 24/25.  
The scanning FPI was mounted inside the SCORPIO-2 multi-mode focal reducer \citep{scorpio2}. 
The operating spectral range around the
[\ion{N}{II}]$\lambda6584$ emission line was cut by a narrow bandpass filter with a $FWHM\approx21\,$\AA\,
bandwidth. The interferometer provides
a free spectral range between the neighbouring interference orders $\sim 35\,$\AA\, with a FWHM of 
the instrumental profile $\sim1.7\,$\AA~ (R$\sim$3860). During the scanning
process, we have consecutively obtained 40 interferograms, each of 1800 s exposure, at different distances between the FPI 
plates, under seeing conditions of $1.7-2.1$\arcsec.  The FoV was $6\farcm1\times6\farcm1$ 
with a sampling of  $0\farcs7$ per pixel. The   data were reduced using   algorithms and programs 
described by \cite{MoiseevEgorov2008} and \cite{Moiseev2015}.  Thus, each spaxel  in the  reduced data 
cube  contains a 40-channel spectrum.

\section{Analysis}

\subsection{Continuum Subtraction}

\begin{figure}
  \centering
     \includegraphics[width=\columnwidth,height=!]{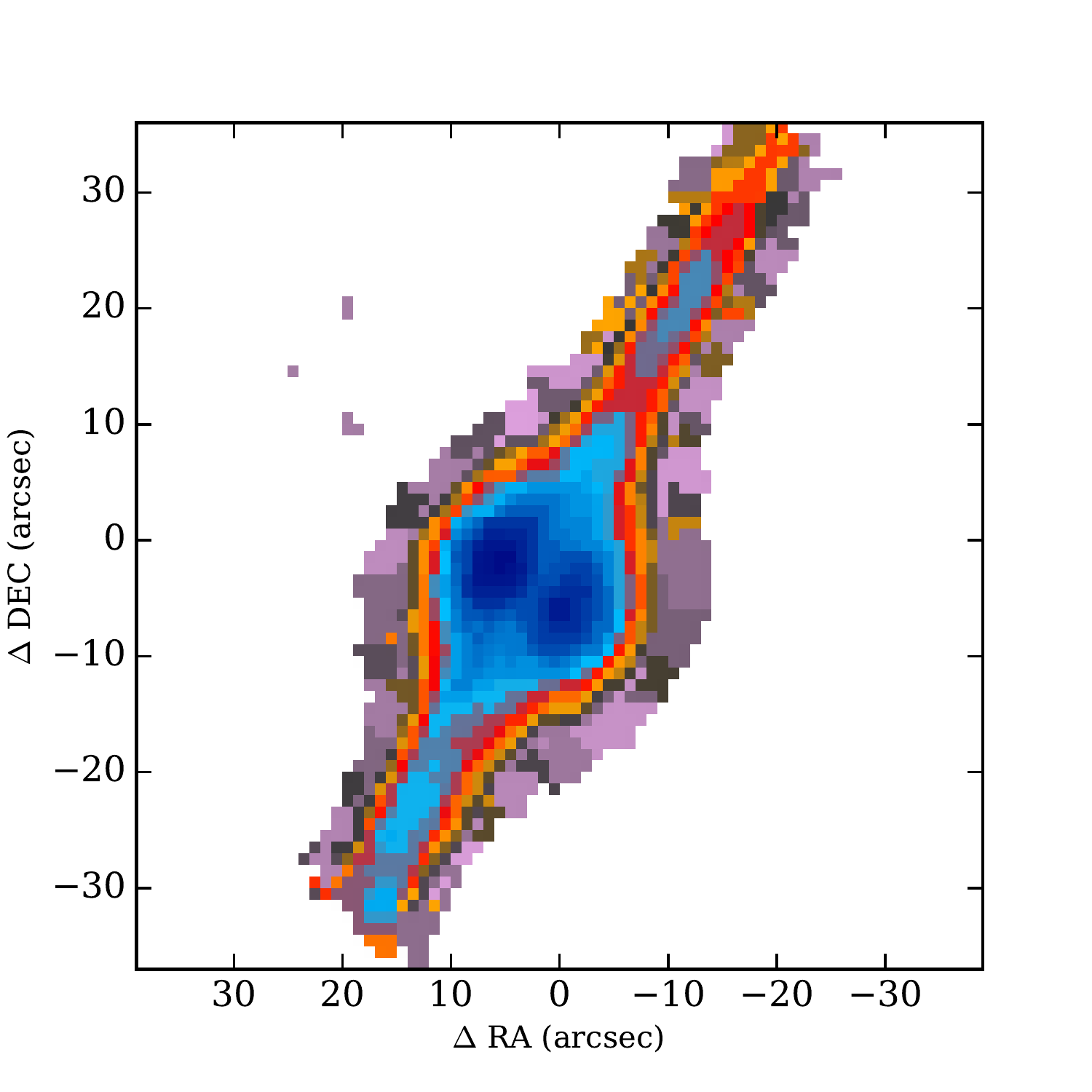}%
    \caption{ Tessellation pattern performed for the analysis of stellar population. 
    The adopted binning procedure ensures that the spatial bins follow the shape of the galaxy. }
    \label{tessellation}
\end{figure}

\begin{figure*}
  \centering
     \includegraphics[clip=true,trim=5cm 0cm 4cm 0cm, width=\textwidth,height=!]{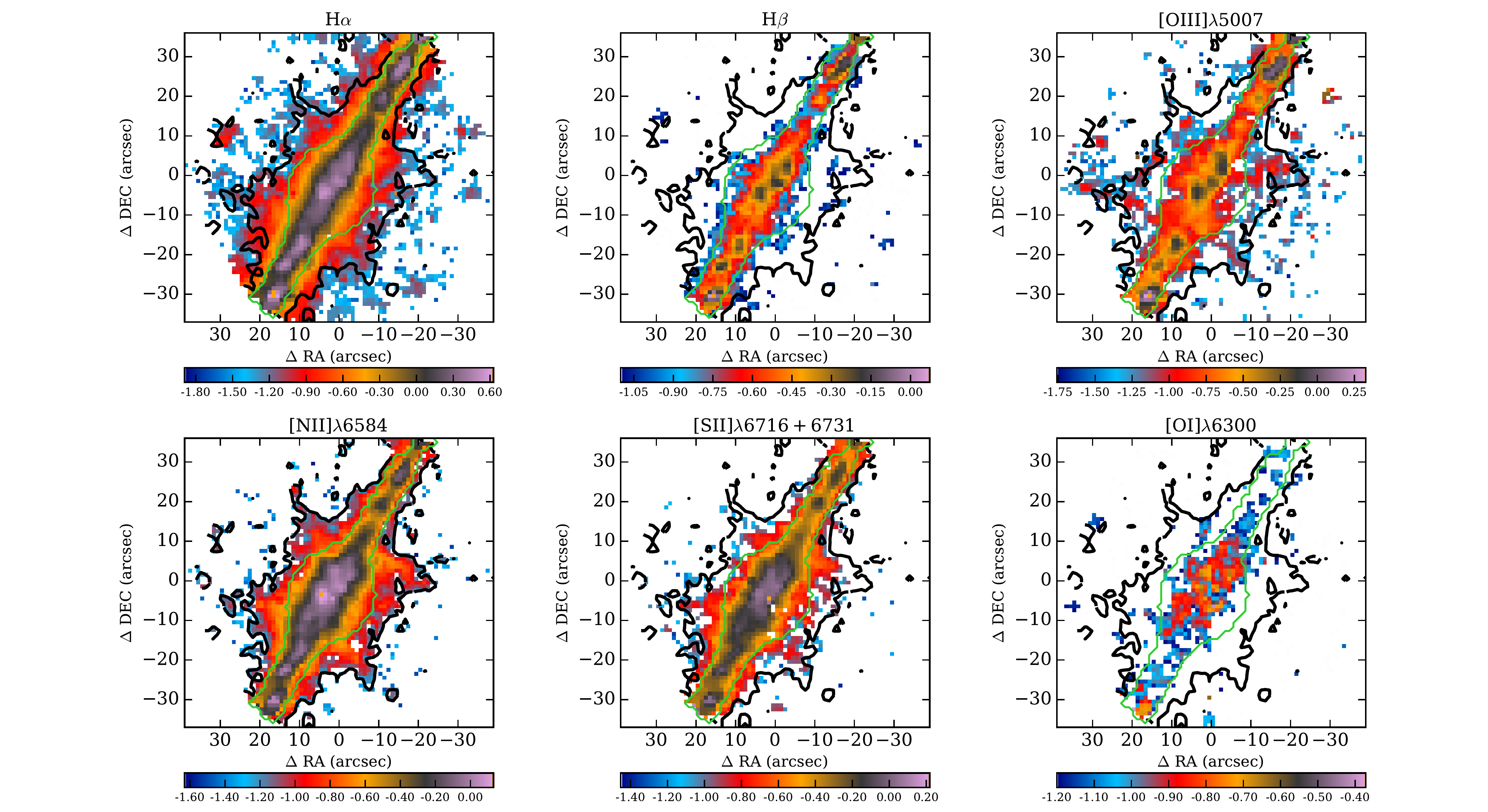}%
  \caption{Example of emission line intensity maps in log scale. The green contour 
  encloses  70\,$\%$ of the flux in the V-band and the black contour is the 
  detection limit of the Nitrogen emission (i.e. for those spaxels with S/N $>$ 5, log\NII = 1.25).
  These two contours demarcate the emission of the disk $+$ bulge and the extraplanar emission.
  The colour bars represent the intensity in log scale of the emission in units of 
   $10^{-16}\,\mathrm{erg\,s^{-1}\,cm^{-2}\,\AA^{-1}}$.}
  \label{mapas_gas} 
\end{figure*}

To study the properties of the ionized gas we need to uncouple the continuum
from each spectra of the data-cube.
There are several routines focused in the analysis of the stellar
populations \citep[e.g. {\sc STARLIGHT}, {\sc
    pPXF},][]{starlight,ppxf}. Here we adopted {\sc Pipe3D}, a
pipeline developed to analyse IFS data-cubes \citep[ S16 hereafter]{fit3d} 
using the fitting package FIT3D \citep[ S15 hereafter]{PIPE3D}.

Generally the continuum spectra show a wide range of signal
to noise ratios (S$/$N) that are higher in the central regions and
gradually decreasing outward from the galactic centre.
From simulations tested in S15, we observed that a minimum \SN~ is
required in each spectrum to obtain an accurate model of the continuum and
therefore to derive the properties of the ionized gas.  In order to
increase the S$/$N in each spectrum we perform a spatial binning
of the cube in the optical range selecting as a goal value a \SN~$\sim$
50. In Fig. \ref{tessellation} we show the result of this tessellation procedure. This tessellation has one advantage compared to other
methods, such as the Voronoi one \citep{voronoi}, that the performed segmentation
follows the morphology of the galaxy and at the same time increases
the \SN, as described in S16.

Once the tessellation is performed, all the spectra within each
spatial bin are co-added and treated as a single spectrum. Then, we derive
a stellar population model for each of those spectra by
performing a multi Synthetic Stellar Population (SSP) fit. Finally, we
recover a model of the stellar continuum for each spaxel by re-scaling
the model within each spatial bin to the continuum flux intensity in
the corresponding spaxel. The stellar population model is then
subtracted to create a pure gas data-cube comprising only the ionized
emission lines (also including the noise and the residuals of the stellar population
modelling). The strongest emission lines within the considered
wavelength range are fitted spaxel by spaxel for the pure gas cube,
adopting a single Gaussian function for each considered emission
line. By doing so, we derive their corresponding flux intensity and the
propagated errors. Finally, we can extract a set of 2D maps comprising
the spatial distribution of the flux intensities for each analysed
emission line. In Figure \ref{ajuste}, we show the results of the fitting
procedure for one single spectrum to illustrate the full process.

\subsection{Ionized gas emission}

In Figure~\ref{mapas_gas} we show the intensity maps for the
emission lines H$\alpha$,
[\ion{N}{II}]$\lambda$6584, [\ion{O}{III}]$\lambda$5007, H$\beta$,
[\ion{S}{II}]$\lambda\lambda$6731,6716 and \OI $\lambda$6300 with a cut in \SN~$>3$. This cut decreases the number
of useful spaxels in each map but not significantly.
From these maps we observe that most of the contribution to the ionized gas is located in a narrow region associated to the disk. There is another evident component of ionized gas that is uncoupled from the disk, located in the extraplanar regions. This extraplanar emission of ionized gas extends perpendicular to the disk and is  
clearly visible in  \Ha, \SII, and \NII~ maps.

\subsection{Spatial variation of the emission line ratios}
\begin{figure*}
  \centering
  \includegraphics[clip=true,trim=5cm 0cm 5cm 0cm, width=\textwidth,height=!]{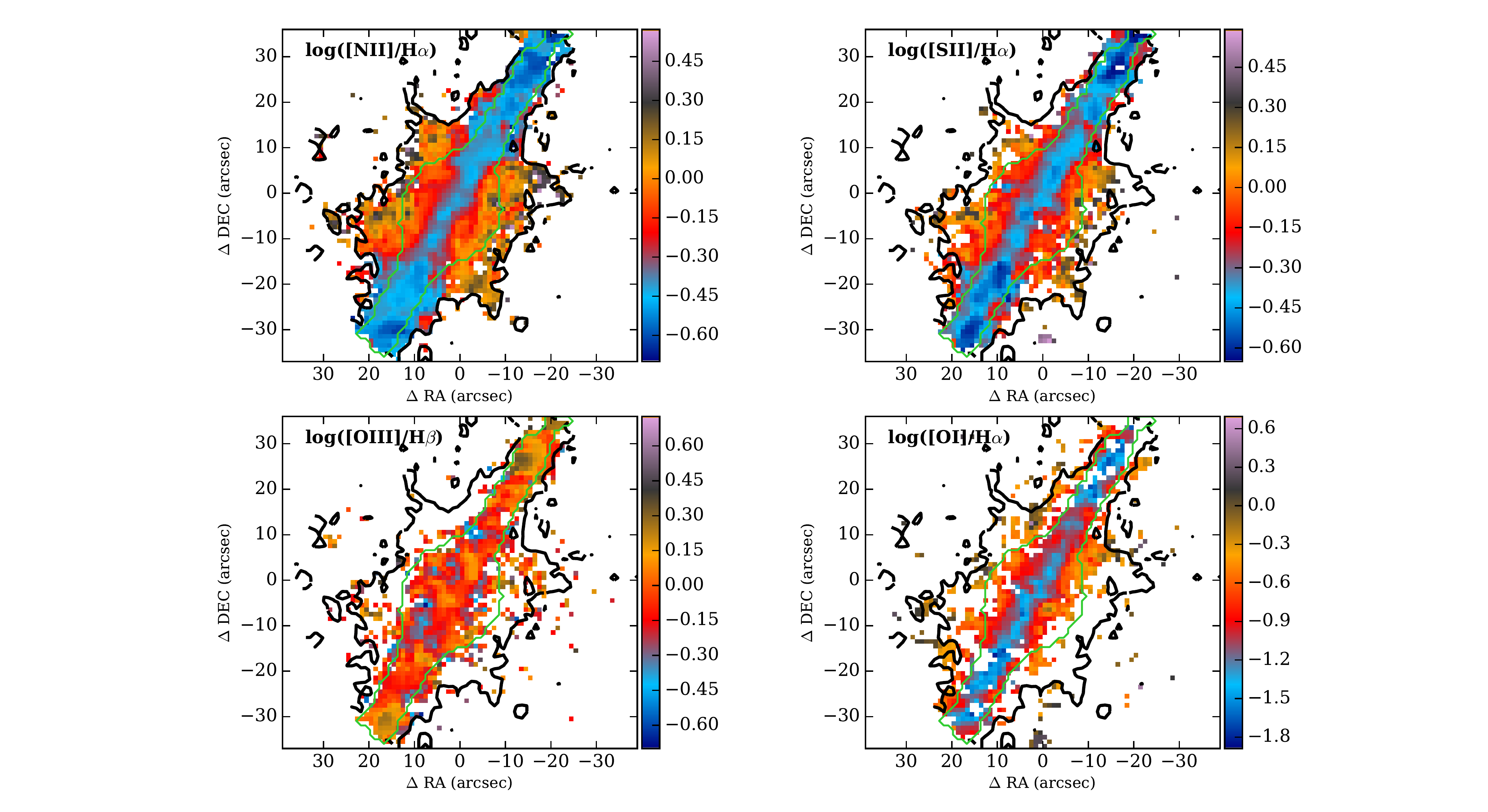}
 \caption{2D line ratio maps for [\ion{N}{II}]/H$\alpha$,
[\ion{S}{II}]/H$\alpha$, [\ion{O}{III}]/H$\alpha$ , and [\ion{O}{I}]/H$\alpha$. [\ion{S}{II}]
represent the sum of the doublet Sulfur ions [\ion{S}{II}] = [\ion{S}{II}]$\lambda$6731 + [\ion{S}{II}]$\lambda$6717. 
The green contour encloses  70\,$\%$ of the flux in the V-band. The black contour represents the same detection limit shown
in Fig.~\ref{rgb}, illustrating the bi-conic structure of the outflow in the [\ion{N}{II}] map.}
  \label{rat_lines}
\end{figure*}

The ionization stage of a photoionized nebula is mainly determined by three parameters: the ionization 
parameter $U$\footnote{$U = Q_0 /4 \pi r^2 N_e c $, where $Q_0$ is the number of ionizing photons emitted 
by the source per second, $r$ is the distance between the source and the nebula, $N_e$ is the electron 
density and $c$ the light speed.}, the hardness of the ionizing radiation (depending on the effective 
temperature and metallicity of the star, or the age and metallicity of the cluster, or the characteristics 
of the AGN), and the optical depth at the Lyman continuum (matter bounded regions having higher ionization stages). 
Furthermore, the ionization stage of shocked regions is determined by the characteristics of the 
shock (speed, magnetic field, densities).


The emission line ratios involving the strong lines \NII, \SII, \OIII, \OI\ and the Balmer lines 
(\Ha\ and \Hb) have been proposed as a method to classify between regions ionized by AGN and
softer ionizing sources as \HII\ regions
\citep[see][]{Veilleux1987}. With the use of IFS, we are able to
spatially resolve the location of different ionization sources by
analysing their line ratios.  As a first
step, we study the spatial distribution of the strongest emission line
ratios and investigate if there is any correspondence with the population
properties.

\begin{figure}
  \centering
  \includegraphics[clip=true,trim=0cm 0cm 0cm 0cm,width=\columnwidth,height=! ]{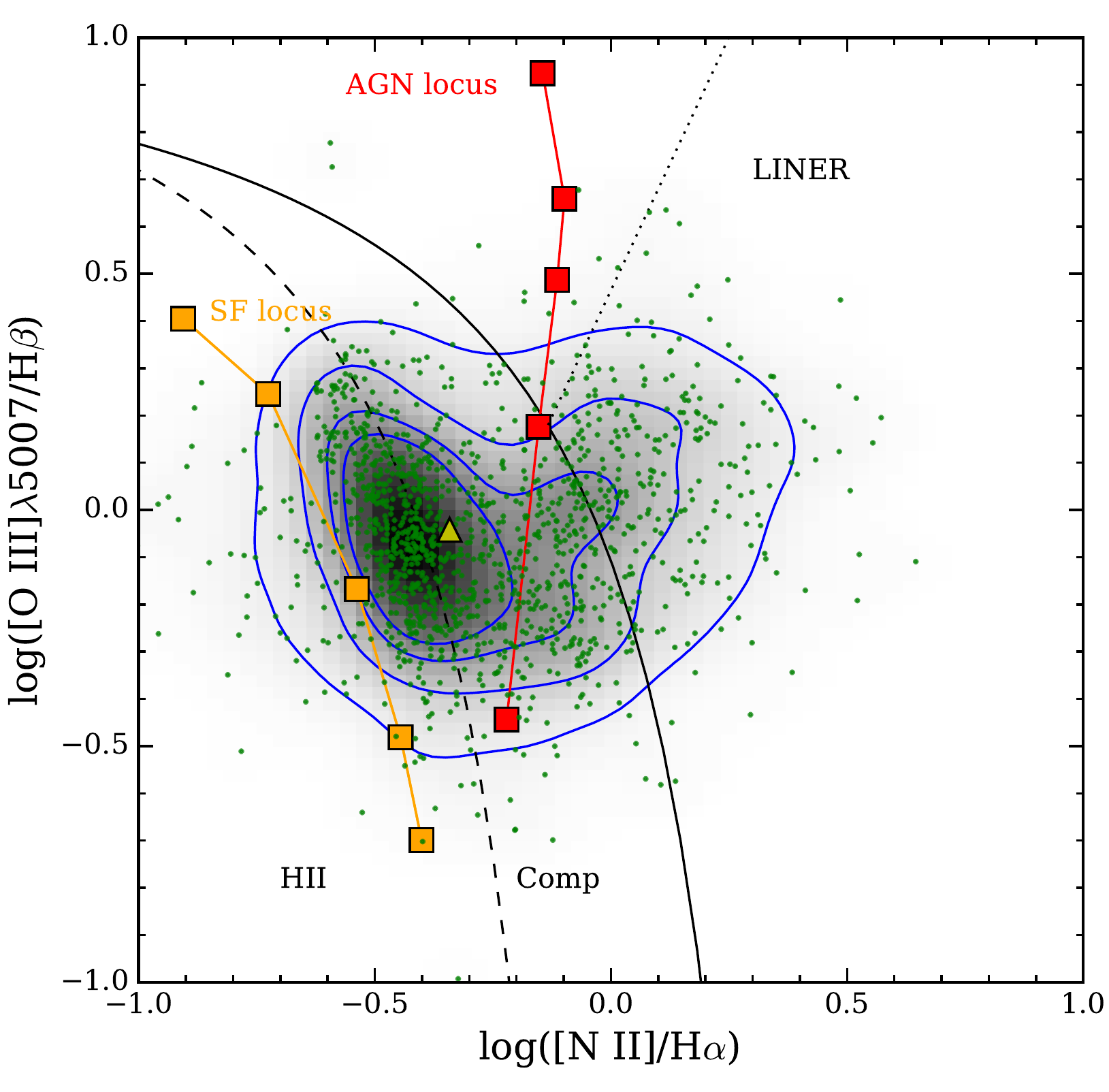}
  \caption{BPT diagnostic diagram ([\ion{N}{II}]/H$\alpha$ versus [\ion{O}{III}]/H$\beta$). Each green dot 
  corresponds to the line intensity ratio at different locations within the galaxy as shown in Fig. \ref{rat_lines}.
  Blue contours are density profiles of the data.
  The largest contour encloses  90$\%$ of the data, while the other ones
  enclose 75, 55 and 35 percent of the total data, respectively. The dashed 
  and solid black lines represent the \citet{kauffmann} and 
  \citet{Kewley2001} curves, respectively. The black dotted line represent the demarcation line from LINER and 
  Seyfert from \citep{kauffmann}
  The yellow triangle is the location of the central spectrum
  of UGC 10043 from the fitting procedure shown in Fig. \ref{ajuste}.
  The red and yellow lines represent the locus of AGN and SF respectively, 
  resulting from reconstructed emission line fluxes based on mean field independent analysis (MFICA) 
  from \citet{AGNloc} and \citet{SFloc}.
  The grey code represents the density of points. We note a 
  bifurcation of the points from the region classically associated with ionization by stars toward regions
  classically associated with AGN ionization.
}
 \label{bpt}
\end{figure}

Figure \ref{rat_lines} shows line ratio maps 
of [\ion{N}{II}]/H$\alpha$, [\ion{S}{II}]/H$\alpha$, 
[\ion{O}{III}]/H$\beta$ and [\ion{O}{I}]/H$\alpha$.
The line ratios  \NII/\Ha\ and \OIII/\Hb\ are insensitive to dust extinction, 
as the emission
lines are close in wavelength. The other emission line ratios are also close
in wavelength, excluding \OI/\Ha.
We include the \OI/\Ha\ emission line in our analysis despite the weakness
of \OI, because
it is a good discriminator between photoionization by hard ionizing power-law
sources and OB stars, and also
because it is a good tracer of shock excitation \citep[e.g.][]{Rich2010,Farage}. 

 A first inspection of the line ratio maps reveals a remarkable increase 
in all line ratios from the
disk toward the extraplanar regions, being more evident in 
the \NII/\Ha\ and \SII/\Ha\ maps. If we assume that \Ha\
traces the extension of the disk as seen in Fig.~\ref{mapas_gas}, then we observe that
the line ratios change  starts at the edge of the disk.

The \OIII/\Hb\ line ratio covers a smaller area due to our cut in \SN\
in \Hb, since \Hb~ is $\sim$3 times weaker than \Ha. On the other
hand, the \OI/\Ha~map is less populated because \OI\ is a weak line
compared to \NII\ and \SII.  The line ratios at the disk region in the
\NII/\Ha\ map are consistent with the maximum ratio (log \NII/\Ha\ $<$
$-0.25$) observed in star forming regions \citep[e.g.][]{kauffmann}. The \OIII/\Hb\ line ratio is more or less constant throughout the galaxy.
The increase in the line ratios \NII/\Ha, \SII/\Ha\ and \OI/\Ha\ far
away from the galactic disk has also been observed in the case of extraplanar 
diffuse ionized gas (DIG) in edge-on galaxies, in which the ionization at 
kiloparsec scales  is produced by hot old stars in the halo with
electron densities in the extraplanar region lower than 10$^{-1}$ cm$^{-3}$  \citep[e.g.][]{Tullmann2000,Flores}.

\subsection{Diagnostic Diagrams}

In order to understand  changes in  line ratios across the field
of view, we explore the so-called diagnostic diagrams.
Baldwin, Phillips and Terlevich proposed a diagram that compares the line 
intensity ratios [\ion{O}{III}]$\lambda5007$/H$\beta$
versus [\ion{N}{II}]$\lambda6583$/H$\alpha$ \citep[known as the BPT diagram, ][]{BPT1981}, to separate the emission
from soft ionizing sources, like \HII~regions, and objects with a higher ionizing power, as
AGNs. However, these line ratios are less accurate to distinguish
among objects of low ionization, like weak AGNs, excitation by shocks, planetary nebulae and
ionization by post--AGB stars \citep[e.g.][]{Binette1994,morisset2009,Binette2009,Cid2011,Kehrig,Singh}.
Subsequently, \citet{Veilleux1987} extended this scheme incorporating
the diagnostic diagrams  for the 
[\ion{S}{II}]/H$\alpha$ and [\ion{O}{I}]$\lambda$6300/H$\alpha$ line ratios. Both line ratios are sensitive to ionization by shocks, being [\ion{O}{I}]/H$\alpha$ the most sensitive. 

Different demarcation lines have been proposed for these three
diagnostic diagrams. The most frequently used ones are those by
\citet{Kewley2001} and \citet{kauffmann}. The most recent ones \citet{AGNloc} and \citet{SFloc} are 
shown in Fig. ~\ref{bpt}. The \citet{Kewley2001} curve
was determined by means of photoionization models. It is the maximum
envelope that can be reached if the ionization is produced by a single
or multiple bursts of star formation.  The \citet{kauffmann} curve was
obtained empirically from galaxies of the Sloan Digital Sky Survey
(SDSS). It traces the approximate upper limit of the sequence of the SF
region in this diagram. The two curves are frequently used to
distinguish from SF regions (below the Kauffmann curve) and
AGNs (above the Kewley curve). The intermediate region between the two
curves is usually interpreted as a composite zone due to a combination of different ionization
sources. However, this can also be populated by ionization due to starburst
galaxies with continuous SF \citep[e.g.][]{Kewley2001},
post-AGB stars \citep[see][]{Cid2011,Papaderos2013, Sanchez2014}, shock ionization \citep[e.g.][]{Rich2010,Ho2014,Alatalo2016}, 
or even classical \ion{H}{ii} regions in evolved areas of galaxies \citep[e.g.][]{Oey1993,Sanchez2015a}. 

In Fig.~\ref{bpt}, we show the classical BPT diagram for the
individual spaxels.  According to the spatial distribution of line ratios
observed in Fig.~\ref{rat_lines}, our interpretation is that the extraplanar
emission is not likely  due to photoionization by young stars.  These points are spread over the region dominated by star
formation towards the region classically dominated by
AGN. So far there is no evidence of the presence of an AGN in 
this galaxy, which is also consistent with its small mass and bulge.
Therefore, we discard AGN as the possible ionization source for UGC 10043.

\begin{figure*}
  \centering
    \includegraphics[clip=true,trim=3cm 4cm 1cm 5cm,width=\columnwidth ,width=\textwidth,height=!]{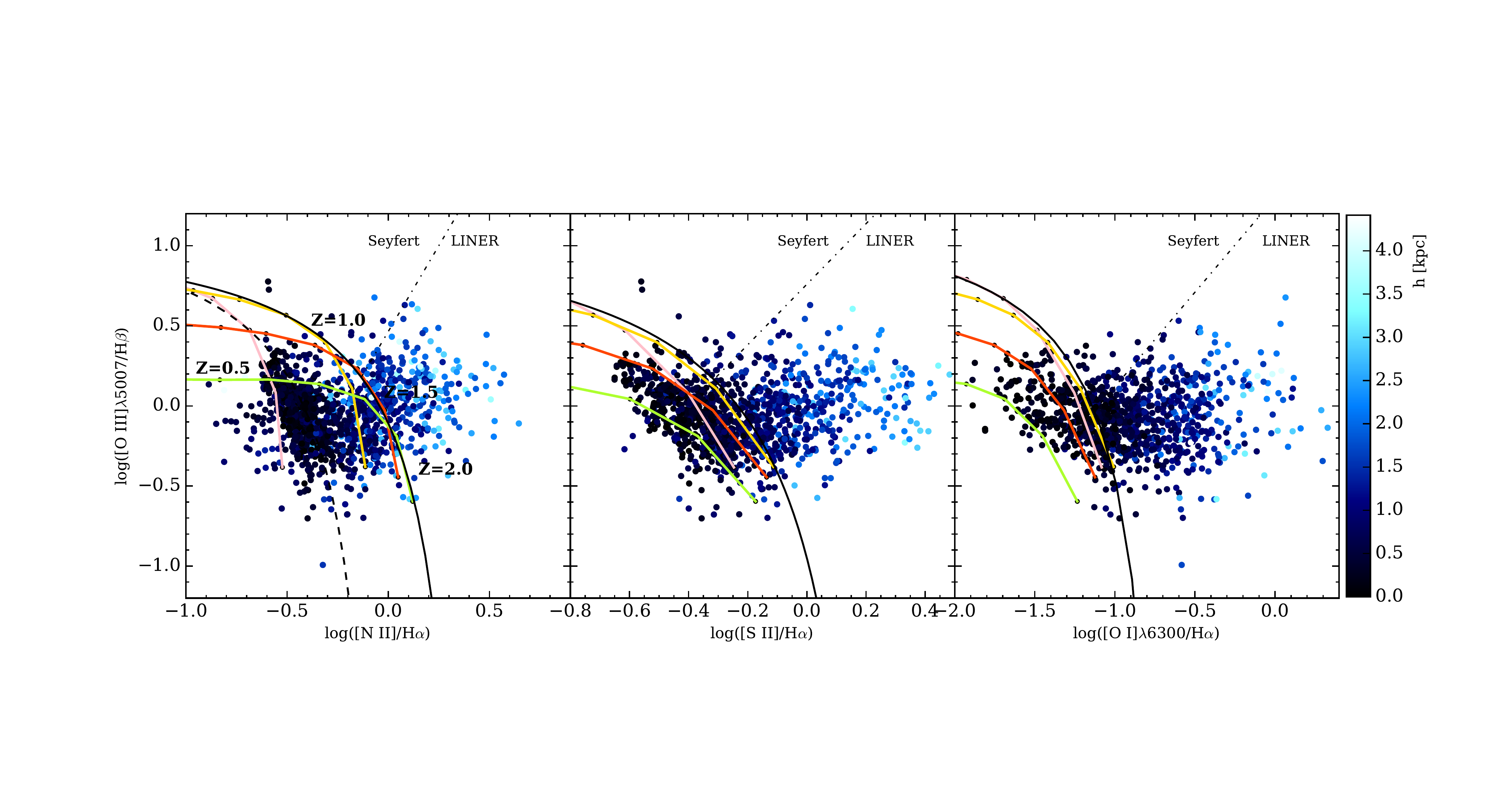}%
  \caption{Diagnostic diagrams for the [\ion{N}{II}]/H$\alpha$,
[\ion{O}{III}]/H$\alpha$ and [\ion{O}{I}]/H$\alpha$ line ratios as a function of \OIII/\Hb.
Each dot corresponds to one spaxel, the colour code represents the perpendicular
distance to the disk, in both directions, in units of kpc. The darkness blue colours are regions closer to the disk and lighter
are those further away. The colour lines represent the predictions of photoionization models from young
stars with continuous-burst for
the different abundances, indicated in the plots, covering a range from Z = 0.5 to Z = 2, with 
different ionization strength at various locations within each curve. The dashed line in the \NII/\Ha~ diagram is
the \citet{kauffmann}  curve and the solid line in the three panels is the 
\citet{Kewley2001} curve for those line ratios. The dotted lines in the 
\NII/\Ha, \SII/\Ha~ and \OI/\Hb~ diagrams are the demarcation lines for LINER and Seyfert galaxies according to \citet{kauffmann} and \citet{Kewley2006}.
  }
  \label{modelos_fotoionizacion}
\end{figure*}

Due to the shape and geometry of the extraplanar emission, the
distribution of line ratios from the star forming region towards the
intermediate AGN area, and the lack of evidence of an AGN, we consider
that the extraplanar emission is related to shock ionization induced
by a galactic wind created by a central SF event.
From Figs. \ref{mapas_gas} and \ref{rat_lines} we observe a symmetrical distribution of the extraplanar gas with respect to the galactic disk. If we assume that the burst of SF was in the nuclear region of the galaxy, and assuming an ideal biconical distribution of the ionized gas \citep[e.g.][]{Heckman} we can trace the limit of the outflowing region based on the distribution of emission lines. The expected conic structure was already shown in Fig.~\ref{rgb}

We will try to confirm the suspicion of shock ionization by comparing the properties
of the emission lines with the predictions by photoionization and shock
models.

\subsection{Photoionionization models}

In Fig. \ref{modelos_fotoionizacion} we compare the observed line ratios with the predicted ones by 
photoionization models along the diagnostic diagrams described in previous sections:
[\ion{N}{II}]/H$\alpha$, \SII/H$\alpha$,
[\ion{O}{I}]/H$\alpha$ versus [\ion{O}{III}]/H$\beta$.
The points are color--coded by their distance from the galaxy disk. The three panels include the predicted values
from the MAPPINGS-III photoionization models \citep{mappings}, for
continuous  SF bursts, and using the PEGASE \footnote{\url{http://www2.iap.fr/pegase/pegasehr/}} synthetic stellar 
library, as described by \citet{dopita2000} and \citet{Kewley2001}.  Each
line corresponds to a different stellar abundance (Z$/$Z$_{\sun}$ = 0.5,
1, 1.5 and 2). For each of them the ionization changes from high-ionization (top-left) to low-ionization (bottom-right).
 In all cases an average electron density of
$n_\mathrm{e}=350$ cm$^{-3}$ is assumed. 
We would like to stress here that the farther a region is located from
 the disk the stronger the ionization it presents (out to $\sim$ 4~kpc).

Thus using a continuous starburst photoionization model we can
reproduce the line ratios that lie below the \citet{Kewley2001} curves, 
whose spatial location is in the disk of the galaxy.
These regions are most probably associated with star
forming regions as we suspected, nevertheless ionization by shocks cover
part of the star forming region, as we will see later.
Furthermore, there are several line ratios in the diagnostic diagrams that
are not covered within the parameters space of these
photoionization models.

\subsection{Shock models}

In Figs.~\ref{bpt} and \ref{modelos_fotoionizacion}, we described two main
components in the ionization mechanism, one produced by young stars at
the disk and an extraplanar emission that is not
reproduced by the analysed photoionization models. Most probably,
this extraplanar ionization can be produced by radiative shocks during the warm phase (T $\sim$ 10$^4$ K)
of a galactic wind as described above. The observed ionized gas can be located over the walls of
the conic structure or in filamentary structures from the nuclear region. 
Shock ionization can be located in the diagnostic diagrams in areas distributed from the
classical location of \HII~regions towards areas where AGN ionization
is dominant \citep[e.g.][]{davies2014}.

\begin{figure*}
  \centering
  \includegraphics[clip=true,trim=0cm 3cm 0cm 4cm,width=\columnwidth ,width=\textwidth,height=!]{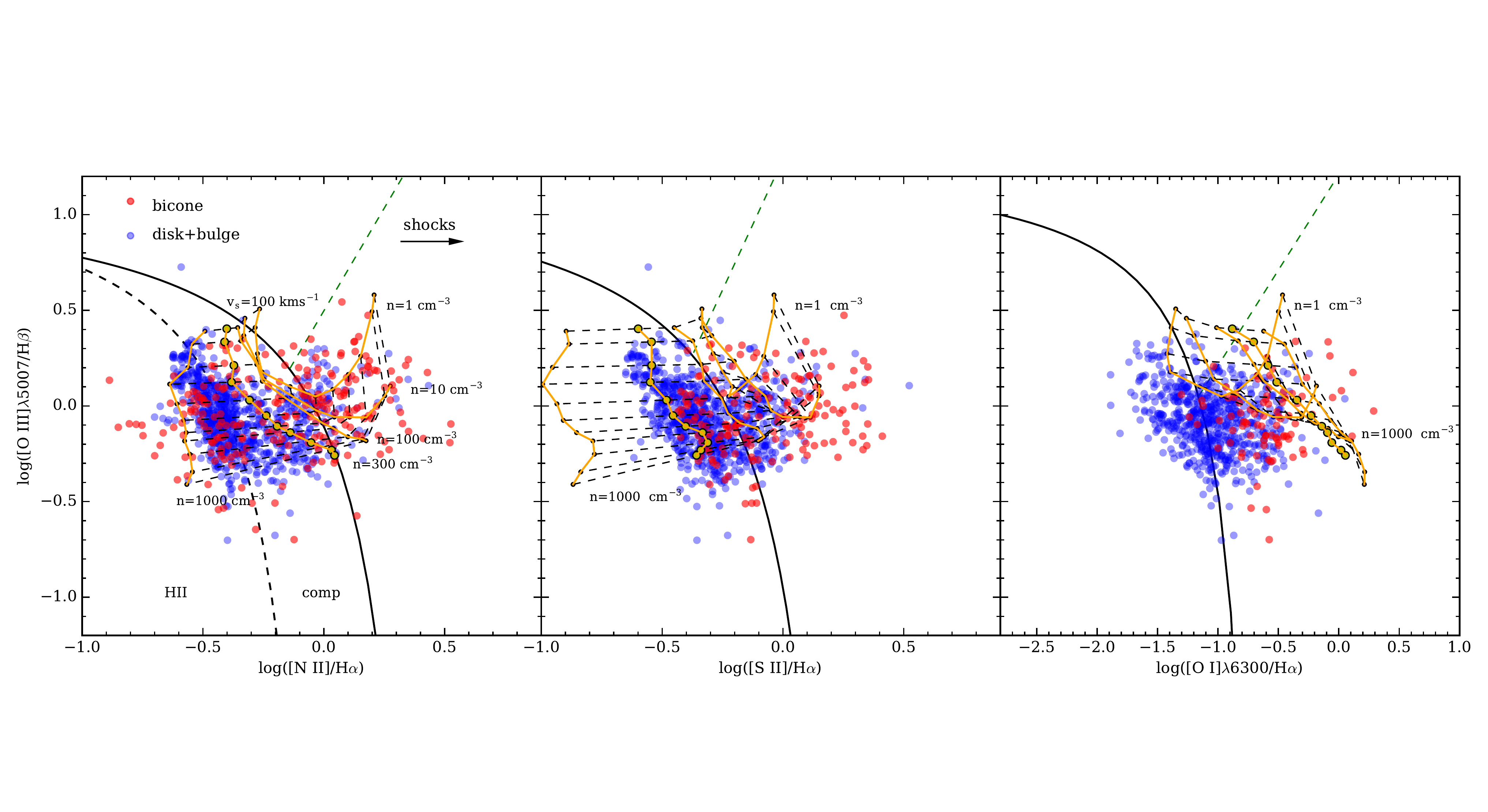}%
  \caption{Diagnostic diagrams for the [\ion{N}{II}]/H$\alpha$,
[\ion{O}{III}]/\Ha and [\ion{O}{I}]/\Ha line ratios as a function of \OIII/\Hb. Red dots
are those located within the bi-conical structure described before and the blue dots are those located in the disk.
Each yellow line represents a shock model with the same magnetic field of B = 5 $\mu$G
and different velocities increasing at the right of each panel from 100 to 400 km s$^{-1}$. The black
dashed lines connect models with equal velocities. Yellow dots are interpolations of the
models of n = 100 cm$^{-3}$ and n = 1000 cm$^{-3}$ at n = 300 cm$^{-3}$.
In each plot, 
we have added a demarcation from two clearest examples of AGN photoinization and shock ionization  
according to \citet{Sharp}. 
This green dashed line marks the bisector between these two fiducial traces, to the left
is  the locus of line ratios for the AGN-excited emission and to the right 
the locus of line ratios for shock-excited emission.
  }
  \label{shock_models}
\end{figure*}

The signatures of shocks can be seen as double peaks or broad components in the
velocity dispersion profiles due to the different kinematic components along
the line of sight and the morphology of the ionized gas.
Due to the low spectral resolution of
the CALIFA data, we are not able to separate the different kinematic
components in the velocity dispersion profiles,
and therefore uncouple and analyse them separately.
At the wavelength of \Ha~ the instrumental
resolution of our data is $\sigma\sim$ 116 km s$^{-1}$, which exceeds the  typical
velocity dispersion of \HII~regions ($<$ 100 km s$^{-1}$, \citet{Yang, MoiseevKlypin2015}). 
Thus, it is not enough to distinguish different kinematic components of the order
expected by shocks \citep[e.g.][]{Lehnert}. Moreover,  even FPI data taken with a 
significant higher spectral resolution ($\sigma\sim$ 33 km s$^{-1}$)
did not resolve a multicomponent  structure of the [\ion{N}{II}] emission line profile,
and  only line broadening and asymmetric profiles were observed (see Sec.~\ref{sec_FPI}).
Broadening profiles in regions outside the galactic disk has also observed in starburst--driven
galactic winds \citep[e.g.][]{Westmoquette}. 
The presence of broad line profiles suggests that complex kinematics components could exist that are
still  unresolved at the spectral resolution of our FP data. 
Therefore, we need to
compare the observed line ratios with the predictions from shock ionization models to determine
if the ionization is driven by shocks, and if so, derive its physical conditions.

We adopt grids of shock models from the MAPPINGS III library of
fast radiative shock models
\citep{Dopita1995,Dopita1996,Dopita2005,mappings}. We selected a shock
velocity ($v_\mathrm{s}$) range of 100 to 400 km s$^{-1}$ in intervals of 25
km s$^{-1}$ and a transverse magnetic field flux density of 5 $\mu$G. For the gas component, we
explored a range including both solar and supra-solar abundances, and
pre-shock densities of 1, 10, 100 and 1000 cm$^{-3}$, respectively.  In
Fig. \ref{shock_models}, we plot the diagnostic diagrams for the
different line ratios explored in Fig. \ref{rat_lines}, together with
the values predicted by the adopted shock models without precursor.
In this plot we have added a different demarcation from the classical LINER--AGN. The new demarcation is a bisector line which separates the ionization by shocks from AGN estimated from two fiducial clear objects of this kind of ionization using IFS data according to \citet{Sharp}.
In addition to the shown models, we have explored many different
combinations of magnetic fields and pre-shock densities, along with
the possible presence of a precursor. However, none of them 
reproduce the observed line ratios in the three diagrams simultaneously,
as well as the  models that are presented in Fig.~\ref{shock_models}.

The shock models in the \NII/\Ha~ and \SII/\Ha~ diagrams cover 
areas frequently associated with continuous SF \citep[e.g.][]{Kewley2001,Sanchez2015a} and 
low ionization nuclear emission--line regions (LINER) \citep[e.g.][]{gomes2016}.
We observe that the majority of the points above the Kewley curves are located
at the right side of the bisector line towards the locus of shock excitation and
therefore in the extraplanar region.
For these diagrams we make
a distinction between the regions that lie in the disk and those in the bi-cones 
according to our demarcation
 shown in Fig \ref{mapas_gas}. There is ionization spatially 
 located at the bi-cones below the Kauffmann curve,
 most probably due to the low shock velocities components in the emission lines.
Our selected shock model describes pretty well the line ratios for the
outflow zone in the \NII/\Ha~ and \SII/\Ha~ diagrams and covers partially the
observed line ratios for  [\ion{O}{I}]/H$\alpha$.
The shock models describe the outflow region for a wind
with a pre-shock density decreasing toward larger extraplanar distances with an increase of 
velocity up to $<$ 400 km s$^{-1}$ in the same direction.

\section{Fabry-P\'erot Interferometer data on gas kinematics}
\label{sec_FPI}

\subsection{Kinematic maps}
\label{sec_kinem}

\begin{figure*}
 \includegraphics[clip=true,trim=0cm 1cm 0cm 1cm,width=\textwidth,height=!]{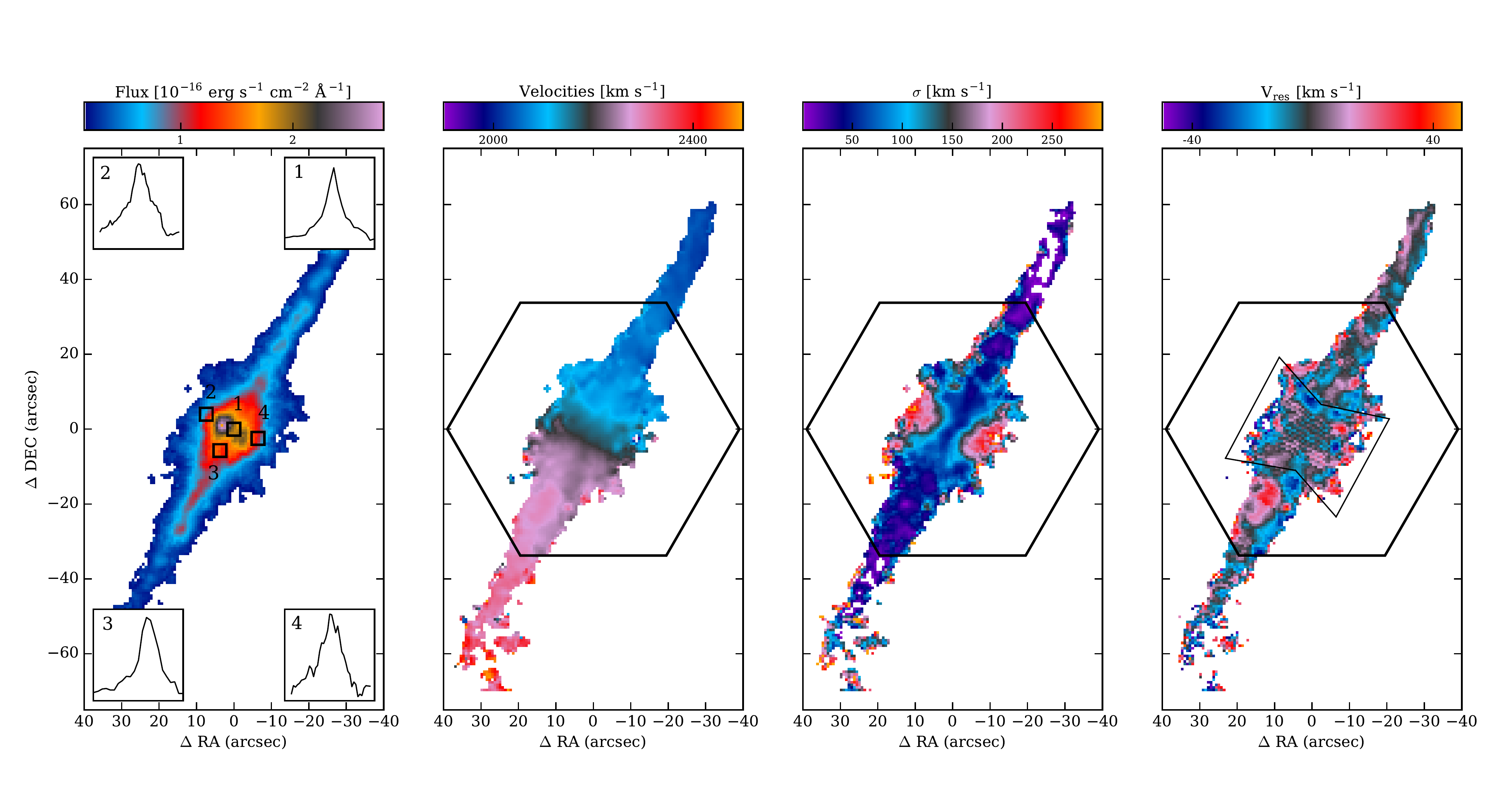}%
  \caption{ Results of the FPI observations in  the [\ion{N}{II}] emission line. 
  From left to the right:   [\ion{N}{II}]  monochromatic map, line-of-sight velocity field, 
  velocity dispersion map free from  instrumental profile influence and map of the
  line-of-sight velocities after subtraction of the mean rotation curve. The insets in the first map are a subset of FPI spectra extracted at different locations. These locations are marked with black squares in the first panel, and they corresponds to apertures of $6.3\arcsec$ $\times$ $6.3\arcsec$}. The hexagon marks
  the field mapped with CALIFA. The geometrical model used for the wind outflow velocities 
  estimation is shown on the last panel.
  \label{FPImaps}
\end{figure*}

The Fabry-P\'erot Interferometer data  provide information  about ionized gas 
kinematics with velocity resolution four times better than that of the PPAK/V500 data, 
in a significantly  larger field-of-view, but with a shallower depth. The observed profiles
of the [\ion{N}{II}]   line were fitted with a Voigt function, which is a convolution
of a Lorentzian and a Gaussian function corresponding to the FPI instrumental profile 
and broadening of observed emission lines, respectively \citep{MoiseevEgorov2008}.  
The emission-line spectrum is very well described by a single-component Voigt profile 
without double or multi-components structures.  Fig.~\ref{FPImaps} shows two-dimensional
maps  derived  from  the fitting process, including the intensity map of   [\ion{N}{II}]$\lambda$6584, 
its line-of-sight velocity field, and its line-of-sight velocity
dispersion ($\sigma$) determined by broadening of emission line. 
The maps are masked using a \SN~$>3$ threshold in the flux intensity. In the first panel
we also plot the spectrum of key regions in the galaxy, showing two regions within the disk, 
and other two in the areas of maximum velocity dispersion located along the semi-minor axis within 
the outflow cone. We observe broad and asymmetric profiles without a remarkable separation of double or
multiple components,
typically found in galactic 
winds driven by SF in these two latter regions \citep[e.g.][]{Eymeren}.

At the outer parts of the  disk we observe a low velocity dispersion (10-40 km s$^{-1}$), which is the typical value 
for giant  \HII~regions in galaxies \citep[e.g.][]{MoiseevKlypin2015}.
While  in the wind bi-cones  $\sigma$ exceeds 100 km s$^{-1}$, increasing towards larger distances from the galactic disk, 
reaching vales $\sim 300$ km s$^{-1}$ probably because at higher distances from the disk the density of the ISM is lower.  These values of the velocity dispersion  are roughly consistent with the shock models 
described before, for which the expected velocities are expected to be lower than 400 km s$^{-1}$ . 

The  FPI map  shows that the circular rotation makes a mayor contribution to the observed 
line-of-site velocities  even in the galactic wind region.
This  picture is typical for edge-on disk galaxies with a relatively moderate 
outflow \citep[e.g.,  NGC\, 4460,][]{OparinMoiseev2015}. 
The mean rotation curve was subtracted from the observed velocity field with the aim to 
remove the regular velocity gradient. We use a model of a transparent rotating cylinder
that provides a good approximation of velocity fields in  edge-on rotating disk galaxies as described in detail in 
\citep{Moiseev2015}. In short, the rotation curve was calculated from averaging  
points across the galaxy major axis, the amplitude of ionized gas rotation  in UGC 10043 
reaches 150 km s$^{-1}$. The mean rotation curve was extrapolated outside the galactic mid-plane
to create a model of  galactic rotation.
The residual line-of-sight velocities after the circular rotation  subtraction are 
shown in the last panel of Fig. \ref{FPImaps}. This map reveals a dozen ``spots'' inside
the outflow bi-cones with typical residual velocities $\pm\,30$ km s$^{-1}$. The   symmetrical 
distribution of the residuals  relative to the major axis implies that we are able to observe real 
regular  deviations  of line-of sight-velocities from the circular rotation produced by 
the wind outflow.

\begin{figure*}
  \includegraphics[width=\textwidth,height=!]{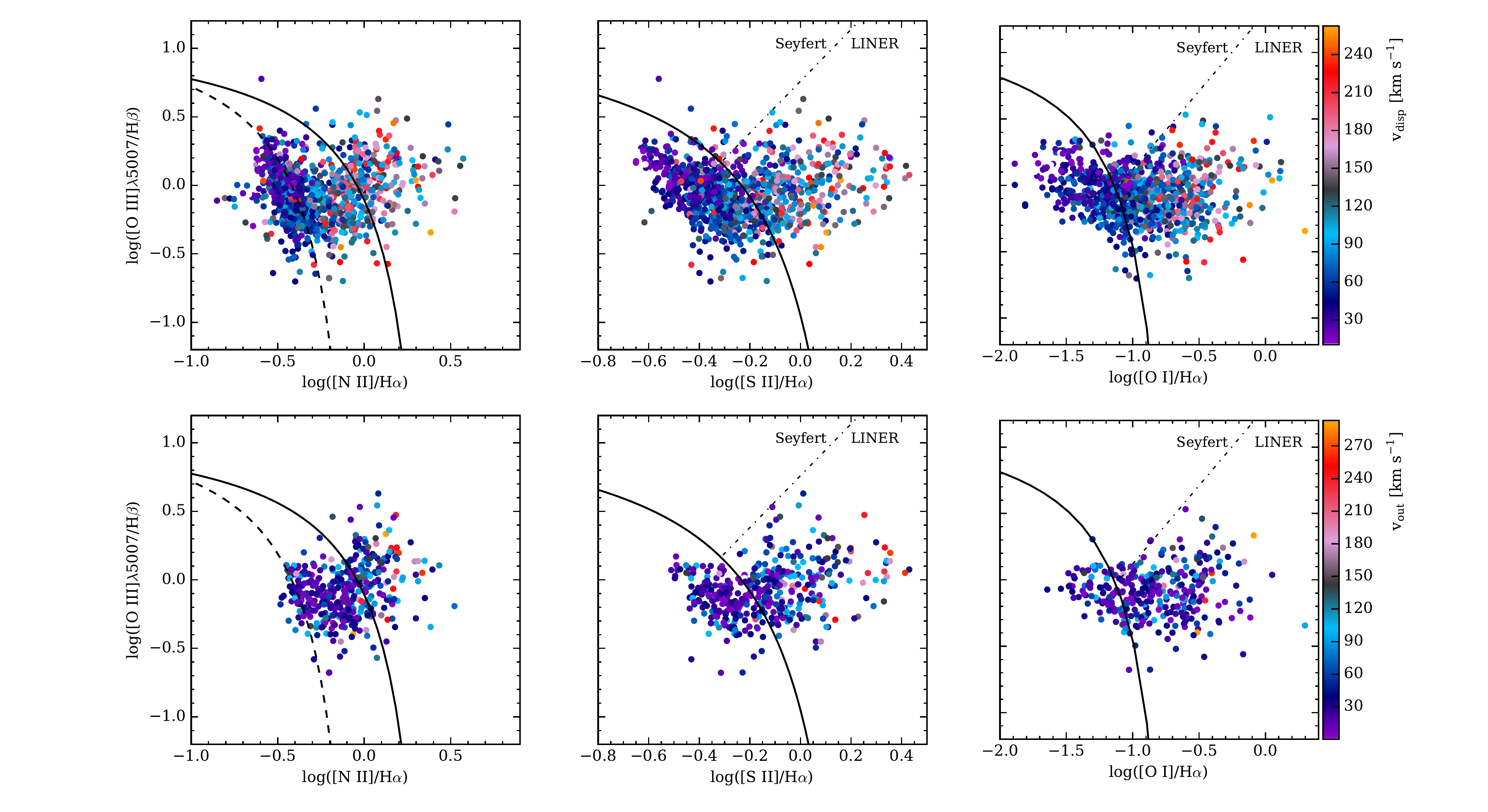}%
  \caption{Diagnostic diagrams similar to Fig.~\ref{modelos_fotoionizacion}, but the
  colour code represents the ionized gas velocity dispersion $\sigma$ (top panels), and $V_{\rm out}$
  (bottom panels) according FPI data. Only points belonging to the bi-cone wind nebulae
  marked on the Fig.~\ref{FPImaps} are shown.}
  \label{FPI_BTP}
\end{figure*}

\subsection{Outflow velocities}

We translated the observed line-of-sight residual velocities into wind outflow
velocities in the frameworks of a simple geometrical model presented by 
\citet{OparinMoiseev2015}. The wind nebulae is  described by frustum rotating 
bi-cones. The matter ejected from the galactic circumnuclear region forms a single 
large shell. While the inner hot gas here is transparent at visible wavelengths, 
the walls of the bi-cones could be observed in optical recombination lines.
From the velocity dispersion map shown in Fig. \ref{FPImaps} it is clear that the shocked ionized gas follows a conical distribution, with a low velocity dispersion at the central regions($<$ 100 km s$^{-1}$),
and at the edges of the cone. Following this kinematic structure we estimate the opening angle as $\theta_{kin}\sim45\degr$ (as marked in 
Fig.~\ref{FPImaps}), traced by eye at the location of the maximum gradient in the velocity dispersion. This aperture angle is smaller than the one one estimated from the morphology of the ionized gas emission based on the CALIFA data. This is most probably because the former traces the location of the observed change in the velocity dispersion, which the latter traces the largest extension of the ionized gas. We note  that according to observations of winds in  other galaxies,
like M 82, cone walls visible in the optical are not homogeneous  but consist of 
several emission filaments.

Under this assumption, the observed gas belongs to  the  cone walls and moves along 
them out of the galactic disk with an outflow velocity $V_{\rm out}$.  For
an edge-on galaxy the positive velocities correspond  to the gas on the walls closest
to us, while the matter with negative velocities moves along the farthest wall.  According
to the formula  in \citet{OparinMoiseev2015} for the galactic inclination $i=90\degr$:
\begin{equation} 
V_{\rm out}=\displaystyle{\frac{V_{\rm res}}{\sin(\theta_{kin}/2)\,\sin\phi}},
\end{equation}
where  $V_{\rm res}$ is the residual line-of-sight velocity after subtracting the pure
circular rotation, $\phi$ is the azimuth angle relative to the axis of the cone. 
With this equation several   regions with a large  amplitude of $V_{\rm res}$ translated
into  fast moving filaments with  $V_{\rm out}\approx$100--250 km s$^{-1}$, while the velocities
of  the neighbouring  points are significantly smaller. These de-projected outflow velocities are in the range of  those found in the surface of bipolar structures \citep[e.g.][]{Heckman,Heckman2003}. Therefore, we confirm the  measurements
of the wind velocities are consistent with shocks models, as in the case of 
the distribution of $\sigma$  (see \S.~\ref{sec_kinem}).

\subsection{Line ratios versus kinematics }
\begin{figure}
\includegraphics[width=\columnwidth, height=!]{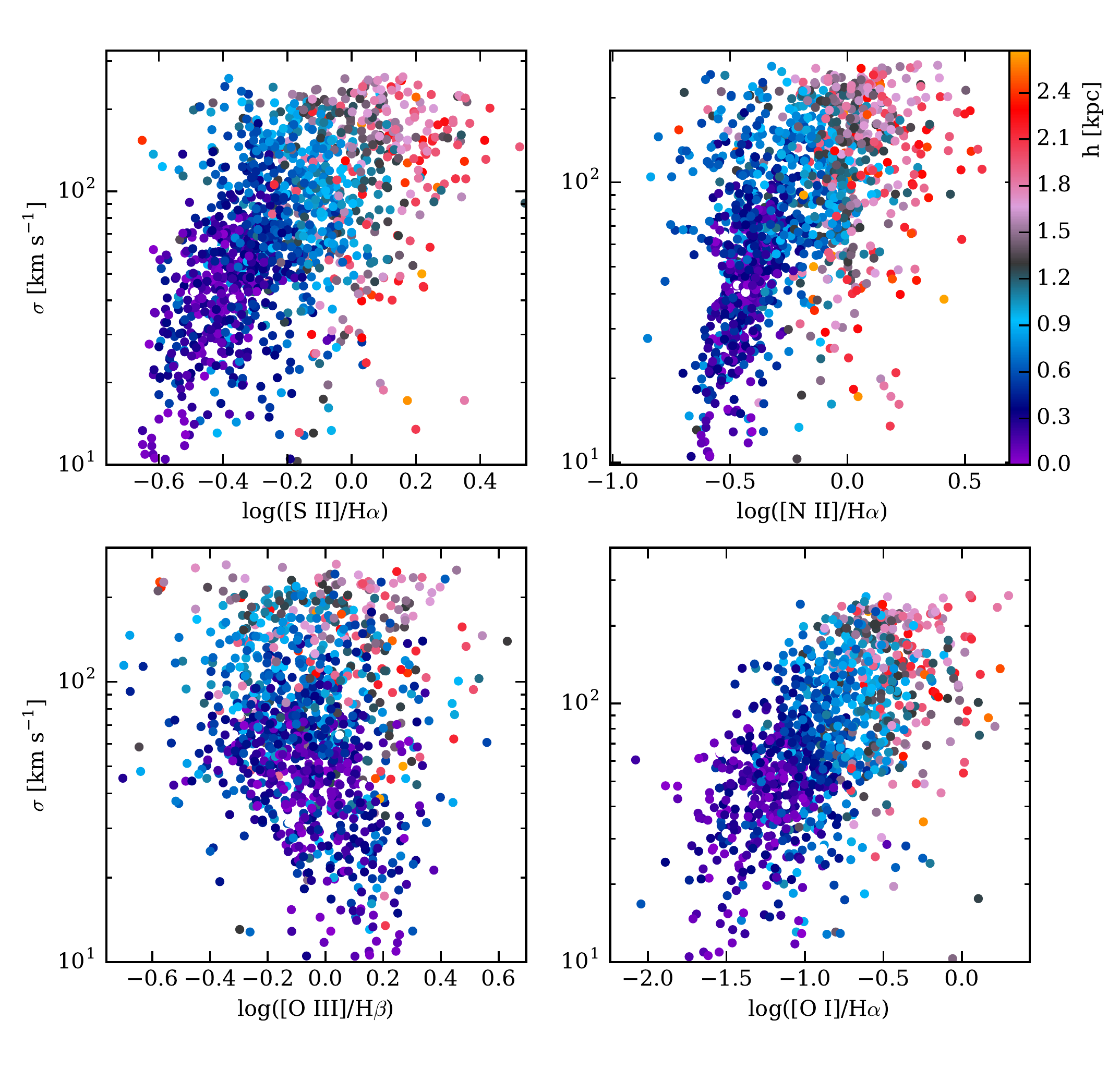}
  \caption{ The correlations between line ratios and ionized gas velocity 
dispersion for the points having both FPI and CALIFA measurements. The colour-bar represent the extraplanar distance in
kpc.}
  \label{sigma_BPT}
\end{figure}

As it was shown before, the kinematic  parameters of the extraplanar outflow derived 
from FPI observations, $\sigma<300$ km s$^{-1}$, and $V_{\rm out}<250$ km s$^{-1}$, are in a good agreement 
with our estimation of shock velocities obtained from the shock models (100--400 km s$^{-1}$). It is 
interesting to compare directly spaxel-to-spaxel  both quantities  with emission line
ratios on BPT diagrams. We re-binned  the FPI maps to the CALIFA pixel scale to make this 
comparison. 

The results are presented in Figs.~\ref{FPI_BTP} and \ref{sigma_BPT}.  Figure~\ref{FPI_BTP} 
show the same diagnostic diagrams shown in Fig.~\ref{modelos_fotoionizacion} and \ref{shock_models}, 
with the values labelled according to the velocity 
and velocity dispersion of the FPI data. The diagrams revealed 
that the points  having high outflow velocity and/or velocity dispersion are indeed located 
in the region corresponded to shock excitation of forbidden lines. This relation is more
clear in the case of velocity dispersion, but the spread of points with the same $V_{\rm out}$
is too high for an unambiguous conclusion. The latter might be related to limitations of 
our very simple geometrical model, as well as with the fact that the points on the diagrams 
do not correspond to these independent points in 
the volume. Indeed, in each pixel  we observe integral values along the line-of-sight for both the lines 
intensities and  their velocities.

We compare the four main emission  line  ratios used throughout  the paper with the corresponding 
velocity dispersion in Fig.~\ref{sigma_BPT}.  This figure  shows a clear  correlation between
the velocity dispersion and the line ratios indicating shock excitation
([\ion{N}{II}]/H$\alpha$, \SII/H$\alpha$, [\ion{O}{I}]/H$\alpha$). While for  the 
[\ion{O}{III}]/H$\beta$   ratio the dependence disappears, because the link  between this  
line ratio and the type  of ionizing source is ambiguous for  [\ion{O}{III}]/H$\beta<10$.
The extraplanar distance reached in these maps is lower than that shown in Fig.~\ref{modelos_fotoionizacion}
due to the limited depth of the FPI data.
Similar plots with a positive correlation between the velocity dispersion  and BPT line 
ratios have been considered by different authors that used 3D-spectroscopic technique  in studies 
of SF galaxies  and ULIRGs \citep[cf.][and references therein]{Monreal2006,Monreal2010, Ho2014, MartinFernandez2016}.  
In the case of UGC\,10043 the correlations shown on the Fig.  \ref{FPI_BTP}, 
and the trend with galactocentric distances, give additional  
arguments in favour of shock-wave origin of the emission lines emission in wind nebulae.

\section{Results}

 Based on our emission maps and the morphology of the outflow we find that
 most probably the wind in UGC 10043 is found during the 
 blow--out phase where the bubble has
 broken. At this stage the wind is free to expand in the intergalactic
 medium carrying metals, mass and momentum.  The [\ion{N}{II}],
 [\ion{S}{II}] and H$\alpha$ intensity maps reveal a gas flux
 following a hourglass shape, originated from the centre of the galaxy
 where the SF is more intense.  Adopting a
 symmetric bi-conic geometry, and following the ionized gas distribution
 from the maps, we determined morphologically the aperture angle of the cones in
 $\theta_\mathrm{morph}\approx$ 80$\degr$ and kinematically in  $\theta_\mathrm{kin}\approx$ 45$\degr$. 
 Here we describe the physical conditions of this outflow.

\subsection{Properties of the ionized gas}
\subsubsection{Electron Density}
\begin{figure}
  \includegraphics[clip=true,trim=0cm 0cm 0cm 2cm,width=\columnwidth ,height=!]{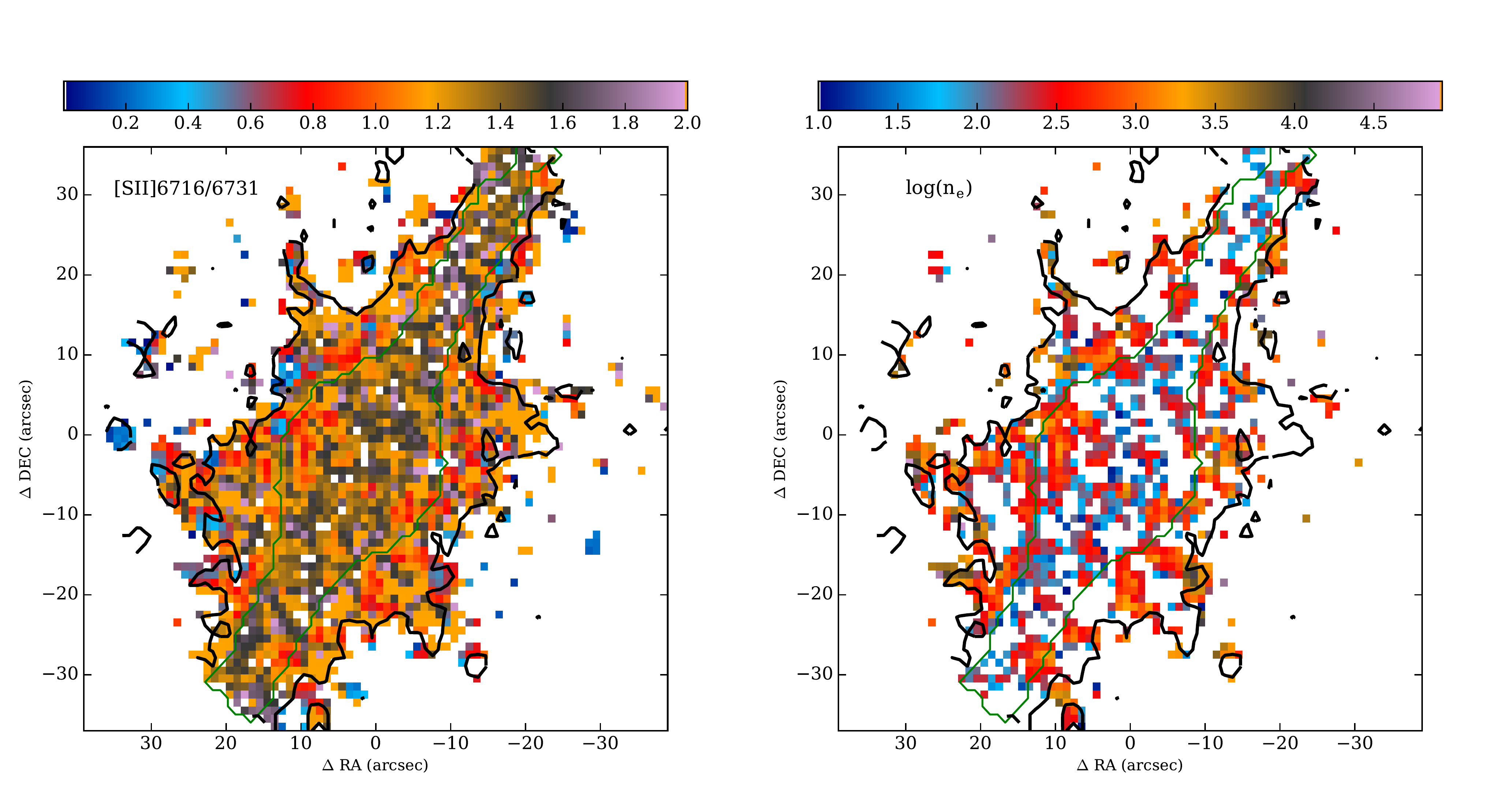}%
  \caption{ Left panel: Spatial distribution of the line ratio $R=$ \SII$\lambda$6716/$\lambda$6731. Right panel: 
  Spatial distribution of the electron density in log scale derived from the 
  \SII~ ratio.
  We show only those regions for which the ratio is sensitive to the electron density, $0.4<R<1.5$.}
  \label{rat_SII}
\end{figure}

From the emission of the Sulfur doublet [\ion{S}{II}]$\lambda
\lambda$6716, 6731 we determine the electron density across the
galaxy \citep[see][]{osterbrock}, by solving the equation
\citep{mccall}:
\begin{equation}
 \frac{I([\ion{S}{II}]\lambda6716)}{I([\ion{S}{II}]\lambda6731)}=\text{1.49}\frac{\text{1+3.77}x}{\text{1+12.8}x}
 \label{SII_eq}
\end{equation}
where $x$ is the density parameter defined as $x=$10$^{-4}n_\mathrm{e}t^{-1/2}$ and $t$ is the 
electron temperature in units of 10$^4$ K. Here we adopted an electron temperature of 10,000 K 
for the optical emission region in shocks, following \citet{Heckman}.
Fig. \ref{rat_SII} shows the spatial distribution of the intensities in the Sulfur ratio together the  electron density derived from Ec. \ref{SII_eq}.
An electron density $n_e\sim$ 200 cm$^{-3}$ is predominant in the whole galaxy, increasing
slightly toward the cones region.
These  electron densities are clearly higher than the expected in DIGs. 
Thus, even in the presence of a DIG, the extraplanar emission is most probably dominated by shocks. 

\subsubsection{H$\alpha$ Luminosity}

Based on the \Ha~flux intensity map, we determine the luminosity
from those regions where the ionization is due to OB stars,
after correcting the flux for dust extinction effects (estimated from the observed \Ha/\Hb~
line ratio), and taking into account the cosmological distance. We
assume the extinction curve of \citet{Cardelli1989} with the specific
attenuation coefficient of $R_\mathrm{v}=$ 3.1 (similar to the
Milky Way one). We assumed an intrinsic line ratio of \Ha/\Hb = 2.86
corresponding to Case B recombination \citep{osterbrock}, for
$T_\mathrm{e}\,=$ 10,000 K and $n_\mathrm{e}\,=$ 100 cm$^{-3}$.  Adopting
a cosmological distance for this galaxy of 35 Mpc, we obtain a total
\Ha~luminosity of (4.47 $\pm$ 0.02)\,$\times\,$10$^{40}$ erg s$^{-1}$ (see Table 1).

\subsubsection{Star Formation Rate}

The \Ha~luminosity is frequently used as an indicator of the Star Formation Rate (SFR) in galaxies
\citep{kennicutt1998}.
After correcting \Ha~ emission by dust extinction it is possible to determine the SFR for those regions
where the ionization is produced by young stars \citep[e.g.][]{catalan}. If we assume that the
regions lying below the Kauffmann curve in the BPT diagram are star forming, and apply 
Kennicutt relation, we can estimate a global SFR = $(0.355 \pm 0.001)$ M$_{\sun}$ yr$^{-1}$.

We also estimate the SFR using the CIGALE code
\citep{Noll}.  This code accounts on the global energy balance between
the energy absorbed by dust in the UV-optical bands and the re-emitted one in the
infrared, and models the spectral energy distributions (SEDs)
as described before. During this procedure the stellar
emission, and both the absorption and emission by dust are taken into
account. The code derives the likelihoods of the current stellar mass and SFR,
from the set of fitted SEDs.
We estimate a SFR of (1.56 $\pm$ 0.23) M$_{\sun}$ yr$^{-1}$.
This value is $\sim 5$ times larger than our previous estimates. This discrepancy in the SFRs is not surprising. The high inclination of UGC 10043 only allows us to observe the emission of the ionized gas located at the most outer parts of the disk. Thus, the observed \Ha~ may not be representative for the overall galaxy emission. The contribution of the ionized gas in the most inner regions are opaque due to the high extinction of the galactic disk. The CIGALE code takes into account both the dust absorption and its re-emission in the infrared, so we expect a higher SFR value.
In the following sections  we will explore the possibility that these SFR estimates are
strong  enough to drive the observed
outflow within the range of the estimated physical parameters.

\subsubsection{Energetic conditions}

Our results indicate that most probably the ionization
is due to shocks by galactic winds driven by recent star formation. 
We would like to explore  the energy injection due to the SFR. 
To do this we follow previous works by 
\citet{Heckman} and \citet{Wild}.

Our data allow us to determine the rate at which the wind carries momentum and energy.
The density in the medium, (i.e. the pre-shock density $n_1$) at which the wind is propagated can be determined through 
the electron density ($n_e$) derived before, if we assume equilibrium between the post--shocked gas and the 
warm gas at T $\sim$ 10$^4$ K:

 \begin{equation}
  n_1[\mathrm{cm^{-3}}]=0.12  \biggl(  \frac{n_\mathrm{e}[\mathrm{cm^{-3}}]}{100}\biggr) \biggl( \frac{350}{v_\mathrm{s}[\mathrm{km\,s^{-1}}]}\biggr)^2 
 \end{equation}
 where $v_s$ is the shock velocity. The thermal pressure
 (P$_\mathrm{gas}$) of the clouds where the line ratios are emitted
 is given by:
 \begin{equation}
   \mathrm{P}_\mathrm{gas}=n_1m_\mathrm{p}\mu v_\mathrm{s}^2=6.63\times10^{-10}\text{Nm$^{-2}$}
 \end{equation}
where $m_\mathrm{p}$ is the proton mass and $\mu$ = 1.36 accounts for an assumed
10$\%$ Helium number fraction and we adopt an electron density $n_e\,\sim\,200$ cm$^{-3}$ (see Sec. 6.1.1).
The momentum flux can be expressed as:
 \begin{equation}
 \mathrm{\dot{p}}_\mathrm{wind}\,=\,\mathrm{P}_\mathrm{wind}(r) \,r^2 \frac{\Omega}{4\pi}  
\end{equation}
where P$_\mathrm{wind}(r)$ is the pressure of the wind at distance $r$. From Fig. \ref{modelos_fotoionizacion}
we estimate a maximum distance of $r\sim$ 4 kpc. The solid angle $\Omega$ is determined
for both apertures angles of the cones 45$\degr$ $<\theta<$80$\degr$.

 In the blow--out phase of a galactic wind the source of pressure can be located in filaments
 originated in the nuclear region or in the external walls of the wind. If the ionized gas
 is emitted in filaments, or in clouds then the pressure of the wind is comparable to the pressure of
 the gas, P$_\mathrm{wind}$ = P$_\mathrm{gas}$. If the ionized gas is located in the walls then the pressure
 of the wind is larger than that of the gas, P$_\mathrm{wind}$ = P$_\mathrm{gas} (v_\mathrm{w}/v_{\mathrm{w}\bot})^2$, where 
 $v_{\mathrm{w}\bot}$ is the perpendicular component of the velocity over the walls. From Fig. \ref{rgb}, 
 $(v_\mathrm{w}/v_{\mathrm{w}\bot})^2=(1/\sin ({\theta}/{2}))^2$.

 Knowing the velocity of the super-wind we could estimate the energy rate:
\begin{equation}
\mathrm{\dot{E}}_\mathrm{wind}=0.5\mathrm{P}_\mathrm{wind}r^2v_\mathrm{w}\frac{\Omega}{4\pi} 
\end{equation}
The velocity of the wind fluid that drives the large scale outflow is poorly constrained in starburst--driven winds. This wind is very hot and tenuous and is quite hard to detect.
Super-wind models predict velocities in the range of
$v_\mathrm{w}$ $\sim$ 1000--3000 km s$^{-1}$ \citep{Hopkins2013}. 
Therefore, if the ionized gas lies in filaments (P$_\mathrm{wind}$ = P$_\mathrm{gas}$ ), then 
we estimate the energy carried out by the wind within this velocity range as:
\begin{equation}
\mathrm{\dot{E}}_\mathrm{wind}(\theta=45\degr)=[0.6-1.8]\times10^{42}\text{erg s$^{-1}$},
\end{equation}
\begin{equation}
\mathrm{\dot{E}}_\mathrm{wind}(\theta=80\degr)=[0.2-0.6]\times10^{42}\text{erg s$^{-1}$}.
\end{equation}
On the other hand, if the ionized gas is located in the walls of the cone, then the energy of the
wind is within the range:
\begin{equation}
\mathrm{\dot{E}}_\mathrm{wind}(\theta=45\degr)=[1.9-5.7]\times10^{42}\text{erg s$^{-1}$} 
\end{equation}
\begin{equation}
\mathrm{\dot{E}}_\mathrm{wind}(\theta=80\degr)=[5.8-17.6]\times10^{42}\text{erg s$^{-1}$} 
\end{equation}

Once estimated the energy of the wind, we can compare it with
the injection rates predicted by supernovae and stellar winds due to SF.
This has been estimated using synthesis models of stellar populations from young stars \citep{Veilleux2005}:
\begin{equation}
 \mathrm{\dot{E}}_*=7\times10^{41}\frac{\text{SFR}}{\text{M}_{\sun}\text{yr}^{-1}}=2.5\times10^{41}\text{ erg s$^{-1}$} 
\end{equation}
While if we use the SFR estimated with CIGALE, then we obtain $ \mathrm{\dot{E}}_*=1.1\times10^{42}$ erg s$^{-1}$. 
We note that in the case of the gas located in the walls the energy of the wind is larger than that due to SF.

\subsubsection{The filling factor}

For the previous calculations we have considered that the whole volume of the cones is occupied 
by gas. 
The ionized gas fraction contained inside the conical 
structure is parametrized by the filling factor $f$. According to \citet{Heckman2000} this factor
is $>$ 10$^{-3}$. The filling factor affects directly in the
derivation of the electron density and therefore the energy of the wind, which has a linear dependence
on the electron density. This means that the real energy of the wind must decrease  up to a factor of 3
orders of magnitude.
This is in concordance with the hypothesis of the ionized gas in the walls, in which the energy from SF
is up to 10$^{-2}$ times the energy of the wind.
Due to the uncertainty in $f$ we can only obtain an upper limit in the energy of the wind, and we are not
able to discard clumpiness or a filamentary structure of the ionized gas.

\section{Conclusions}

We used the unique capabilities of the CALIFA survey to study
the emission of the
ionized gas in the edge on galaxy UGC 10043.  We detect the presence of {a possibly bi-conical}
extraplanar emission of ionized gas that reaches distances up to 4 kpc
over the galaxy disk. {We favour, that} this extraplanar emission is most probably 
produced by a galactic wind driven by a weak nuclear starburst, consistent with a low  SFR = $0.35$ M$_{\sun}$ yr$^{-1}$ as measured with \Ha.

Based on shock models, we find that our data are better described by a fast
velocity shock model ($v_s$ = 100--400 km s$^{-1}$), and a pre-shock density
decreasing towards larger extraplanar regions.  These shock velocities are in  agreement  to first order with
the outflow kinematic parameters derived from scanning Fabry-P\'erot observations: the line-of-sight velocity dispersion 
$\sigma<300$ km s$^{-1}$, and  the de-projected outflow velocity $V_{\rm out}<250$ km s$^{-1}$. 

This study stresses the necessity of exploring these events in a more
systematic way. In particular by using a large sample of galaxies
observed using similar IFU techniques, like the ones provided by surveys
such as CALIFA or MaNGA \citep{manga} which provides a higher spectral resolution.  Only with a large statistical
baseline, we will be able to determine if the apparent deficit of
star formation derived with H$\alpha$ described for this galaxy, when
compared to the required energy to support the outflow, is a general
property or not. Moreover, it is also important to determine if the use of
analysis of the integrated SED of the galaxies including infrared
photometry could overcome this problem in general.

\section*{Acknowledgements}

The authors wish to thank the anonymous referee for his/her thorough review,  and the valuable suggestions and 
comments which significantly contributed to improve the quality of the publication.
CALIFA is the first legacy survey being performed at Calar Alto. The CALIFA collaboration would like to thank 
the IAA-CSIC and MPIA-MPG as major partners of the observatory, and CAHA itself, for the unique access
to telescope time and support in manpower and infrastructures. The CALIFA collaboration thanks also the CAHA 
staff for the dedication to this project. 

We thank CONACYT-125180, DGAPA(UNAM)-IA100815 and DGAPA(UNAM)-IN107215 projects and the 
institutions (CONACYT and DGAPA-UNAM) for providing financial support for this study.

The Fabry-P\'erot observations were obtained   with the 6-m  telescope of 
the Special Astrophysical Observatory of the  Russian Academy of Sciences, and  were  carried out 
with the financial support of  the Ministry of Education and Science of the Russian Federation 
(Contract No. N14.619.21.0004 for the project RFMEFI61914X0004). AVM and DVO are also grateful for
the financial support via grant MD3623.2015.2 from the President of the Russian Federation.
L.G. was supported in part by the US National Science Foundation under Grant AST-1311862.
RAM acknowledges support by the Swiss National Science Foundation.




\bibliographystyle{mnras}
\bibliography{ref} 





\bsp	
\label{lastpage}
\end{document}